\documentclass[aps,prx, reprint,superscriptaddress]{revtex4-1}

\usepackage{amsmath}    
\usepackage{graphicx}     
\usepackage{verbatim}     
\usepackage{color}           
\usepackage{subfigure}    
\usepackage[colorlinks,hyperindex]{hyperref}
	\hypersetup
		{colorlinks,	%
	 	citecolor=blue,	%
	 	linkcolor=blue,	%
	 	urlcolor=blue,	%
		}

\usepackage{datetime}

\usepackage{notes2bib}

\begin{document}


\title{Data-driven studies of magnetic two-dimensional materials} 

\author{Trevor David Rhone}
\email{trr715@g.harvard.edu}
\affiliation{Department of Physics, Harvard University, Cambridge, Massachusetts 02138, USA}

\author{Wei Chen}
\affiliation{Department of Physics, Harvard University, Cambridge, Massachusetts 02138, USA}

\author{Shaan Desai}
\affiliation{Department of Physics, Harvard University, Cambridge, Massachusetts 02138, USA}

\author{Amir Yacoby}
\affiliation{Department of Physics, Harvard University, Cambridge, Massachusetts 02138, USA}
\author{Efthimios Kaxiras}
\affiliation{Department of Physics, Harvard University, Cambridge, Massachusetts 02138, USA}
\affiliation{School of Engineering and Applied Sciences, Harvard University, Cambridge, Massachusetts 02138, USA}

\date{\today}

\begin{abstract}
We use a data-driven approach to study the magnetic and thermodynamic properties of van der Waals (vdW) layered materials.
We investigate monolayers of the form A$_2$B$_2$X$_6$, based on the known material Cr$_2$Ge$_2$Te$_6$, using 
density functional theory (DFT) calculations and machine learning methods to determine their 
magnetic properties, such as magnetic order and magnetic moment. 
We also examine formation energies and use them as a proxy for chemical stability.  
We show that machine learning tools, combined with DFT calculations, can provide a 
computationally efficient means to predict properties of such two-dimensional (2D) magnetic materials. 
Our data analytics approach provides insights into the microscopic origins of magnetic ordering in these systems.
For instance, we find that the X site strongly affects the magnetic coupling between neighboring A sites, which drives the magnetic ordering.
Our approach opens new ways for rapid discovery of chemically stable vdW materials that exhibit magnetic behavior.  
\end{abstract}

\maketitle

\section{Introduction}

The discovery of graphene ushered in a new era of studies of materials properties in the two-dimensional (2D) limit~ \cite{robinson2015}.  
For many years after this discovery only a handful of van der Waals (vdW) materials were extensively studied.
Recently, over a thousand new 2D crystals have been proposed~\cite{cheon2017,Marzari}.
The explosion in the number of known 2D materials increases demands for probing them for exciting new physics 
and potential applications~\cite{chen2013tuning,nourbakhsh2016mos2}.
Several 2D materials have already been shown to exhibit a range of exotic properties including superconductivity, topological insulating behavior
and half-metallicity~\cite{novoselov20162d,zeng2016enhanced,choi2017atomistic,chen2017properties}. 
Consequently, there is a need to develop tools to quickly screen a large number of 2D materials for targeted properties.
Traditional approaches, based on sequential quantum mechanical calculations or experiments are usually slow and costly.
Furthermore, a generic approach to design a crystal structure with the desired properties, although of practical significance, does not exist yet.  
Research towards building structure-property relationships of crystals is in its infancy~\cite{mbtr, riley2017, tropsha2017}. 

Long-range ferromagnetism in 2D crystals has recently been discovered~\cite{huang2017, cava2017}, sparking
a push to understand the properties of these 2D magnetic materials and to discover new ones with improved 
behavior~\cite{cui2017contrasting,miyazato2018, moller2018, sun2018}.
2D crystals provide a unique platform for exploring the microscopic origins of magnetic ordering in reduced dimensions. 
Long-range magnetic order is strongly suppressed in 2D according to the Mermin-Wagner theorem~\cite{mermin1966},
but magnetocrystalline anisotropy can stabilize magnetic ordering~\cite{hope2000}. 
This magnetic anisotropy is driven by spin-orbit coupling
which depends on the relative positions of atoms and their identities.
As a result, the magnetic order should be strongly affected by changes in the structural arrangements of atoms and chemical composition of the crystal.

Chemical instability presents a crucial limitation to the fabrication and use of 2D magnetic materials.  
For instance, black phosphorous degrades upon exposure to air and thus needs to be handled and stored in vacuum or under inert atmosphere. 
Structural stability is a necessary ingredient for industrial scale application of magnetic vdW materials, such as CrI$_3$ and Cr$_2$Ge$_2$Te$_6$~\cite{huang2017,cava2017}.
In addition to designing 2D materials for desirable magnetic properties, 
it is important to screen for those materials that are chemically stable.
In our approach, we employ the calculated formation energy as a proxy for the chemical stability~\cite{rasmussen2015}.
In particular, we obtain the total energies of systems at zero temperature,
and calculate the formation energy as the difference in
total energy between the crystal and its constituent elements in their respective crystal phases. 
This quantity determines 
whether the structure is thermodynamically stable or would decompose. 
This formulation ignores the effects of zero-point vibrational energy and entropy on the stability.

Recently, machine learning (ML) has been combined with traditional methods (experiments and ab-initio calculations) to advance rapid materials discovery~\cite{rupp2012, meredig2014, tsuda2015, rasmussen2015, Ueno2016, Francesca2017, tsuda2017, cheon2017, Marzari}. 
ML models trained on a number of structures can predict the properties of a much larger set of materials. 
In particular, there is presently a growing interest in exploiting ML for discovery of magnetic materials~\cite{landrum2003, moller2018}.  
Data-driven studies of ferromagnetism in transition metal alloys have highlighted the importance of novel data analytics techniques to tackle problems in condensed matter physics~\cite{landrum2003}.
It is conceivable that tuning the atomic composition could provide an additional degree of freedom in the search for stable 2D materials with interesting magnetic properties~\cite{janus2017}.
Even more compelling is the ability of ML tools to assist in uncovering the physics underlying the stability and magnetism of 2D materials~\cite{cubuk2016, cubuk2017}.  
%
Specifically, ML methods can identify
patterns in a high-dimensional space revealing relationships that could be otherwise missed.

\section{Methodology}
%
%

\begin{figure}[h!]
\includegraphics[width=0.48\textwidth,keepaspectratio]{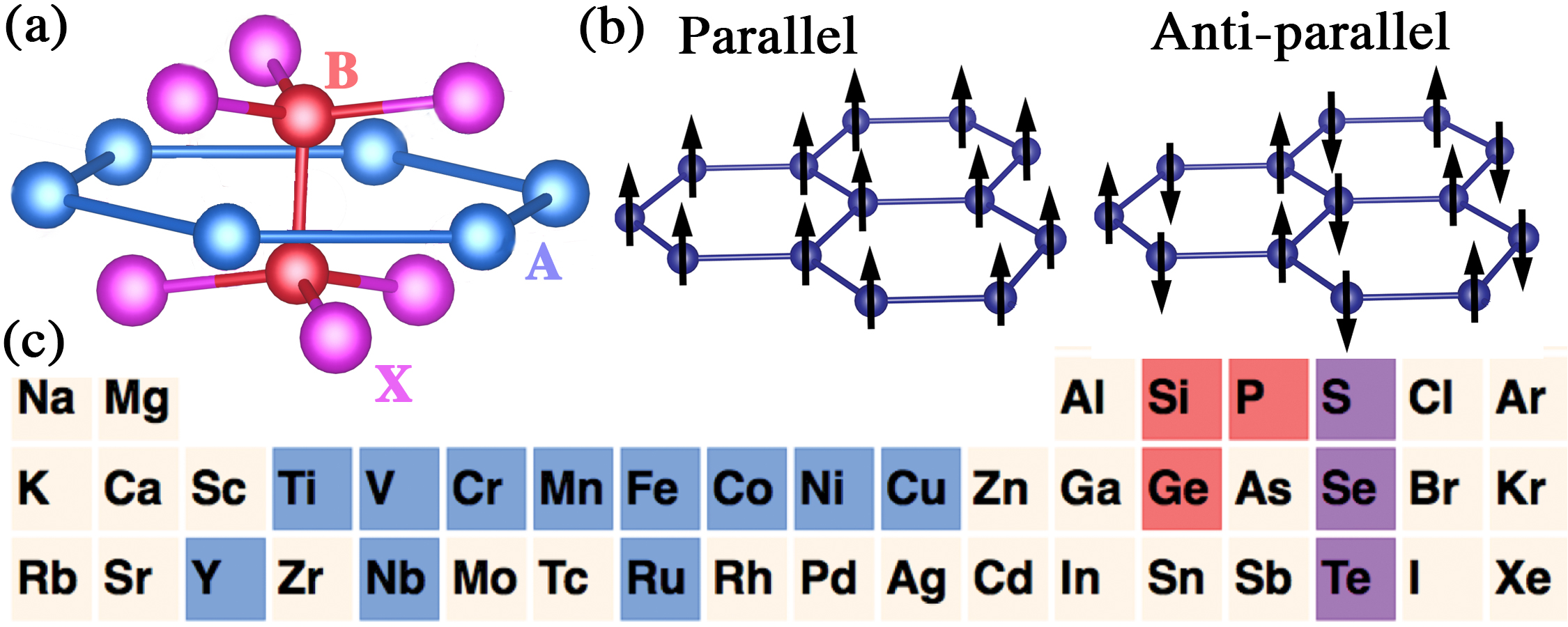}
 \caption{\label{fig1}(a) Crystal structure of the A$_2$B$_2$X$_6$ lattice. (b) Magnetic orders considered in the A plane, labelled parallel and anti-parallel. 
 (c) Elements used for substitution of A (blue), B (red) and X (magenta) sites. 
 }
\end{figure}

In order to develop a path towards discovering 2D magnetic materials, we generate a database
of structures based on a monolayer Cr$_2$Ge$_2$Te$_6$ (Fig.~\ref{fig1}(a)) using density functional theory (DFT) calculations
~\footnote{The results of these DFT calculations will be used to build a database of monolayer 2D materials 
which will be publicly available to the scientific community.}. 
%
The possible structures amount to a 
combinatorially large number of type A$_2$B$_2$X$_6$ ($\sim10^4$) with different elements occupying the A, B and X sites.  
We select a subset of 198 structures due to computational constraints. 
We obtain the total energy, magnetic order, and magnetic moment of each structure. 
The ground-state properties were determined by examining the energies of the fully optimized structure with several spin configurations, including non-spin-polarized, parallel, and anti-parallel spin orientations at the A sites (Fig.~\ref{fig1}(b)).

%

We then employ
a set of materials descriptors which comprise easily attainable atomic properties,
and are suitable for describing magnetic phenomena.
We employ additional descriptors which are related to the formation energy~\cite{Ulrich2004}.
The performance of descriptors in predicting the magnetic properties or thermodynamic stability sheds some light into the origin of these properties.

%
%
To create the database we use
DFT calculations 
~\footnote{We used the GGA-PBE for the exchange-correlation functional. The energy cutoff was 300 eV. 
The vacuum region was thicker than 20 \AA. 
The atoms were fully relaxed until the force on each atom was smaller than 0.01 eV/\AA. 
A $\Gamma$-centered 10$\times$10$\times$1 \emph{k}-point mesh was utilized.} 
with the VASP code~\cite{vasp3}.
%
We create the different structures by 
substituting one of two Cr atoms (A site) in the unit cell with a transition metal atom,
from the list:  
Ti, V, Cr, Mn, Fe, Co, Ni, Cu, Y, Nb, Ru. 
In the two B sites we place combinations of Ge, Si, and P atoms, namely Ge$_2$, GeSi, GeP, 
Si$_2$, SiP, P$_2$.
The atoms at X sites were either S, Se, or Te, that is, S$_6$, Se$_6$, Te$_6$. 
Fig~\ref{fig1}(c) shows the choice of substitution atoms in the Periodic Table.
An example of a structure created through this process is (CrTi)(SiGe)Te$_6$.

%
%
%
%
The careful choice of descriptors is essential for the success of any ML approach~\cite{Scheffler2015, Tanaka2017}. 
We use atomic properties data from the python mendeleev package 0.4.1~\cite{mendeleev2014} to build descriptors for our ML models. 
We performed supervised learning with atomic properties data as inputs, with
target properties the magnetic moment and the formation energy.
%
The choice of the set of descriptors for the magnetic properties was motivated by the Pauli exclusion principle, which gives rise to the exchange and super-exchange interactions. 
We also consider the magneto-crystalline anisotropy~\cite{chikazumi} 
by building inter-atomic distances and electronic orbital information into our descriptors.
%
%
With respect to the formation energy, the choice of descriptors was motivated, in part, by the extended Born-Haber model~\cite{Ulrich2004},
%
and include the dipole polarizability, the ionization energy and the atomic radius (see Supplemental Materials for a full list of atomic properties and descriptors used~\footnote{See Supplemental Material at http://link.aps.org/ supplemental/10.1103/PhysRevX.x.xxxxx for more details and discussions related to this letter. \label{footnote_sm}}).

%
The data were randomly divided into a training set, a cross-validation set and a test set. 
Training data and cross-validation were typically 60\% of the total data while test data comprised 40\% of all the data. 
We employed the following ML models: kernel ridge regression, extra trees regression, and neural networks.  
Kernel ridge regression with a gaussian kernel has been shown to be successful in several materials informatics studies.
Extra trees regression allows us to determine the relative importances of features used in a successful model~\cite{hastie}. 
An analysis of hidden layers of the deep neural networks could allow us to identify patterns in 2D materials properties data,
thereby guiding theoretical studies~\cite{cubuk2017}.


\section{Results and discussion}

\subsection{Magnetic properties}

%
\begin{figure}[h]
  \includegraphics[width=0.48\textwidth, keepaspectratio]{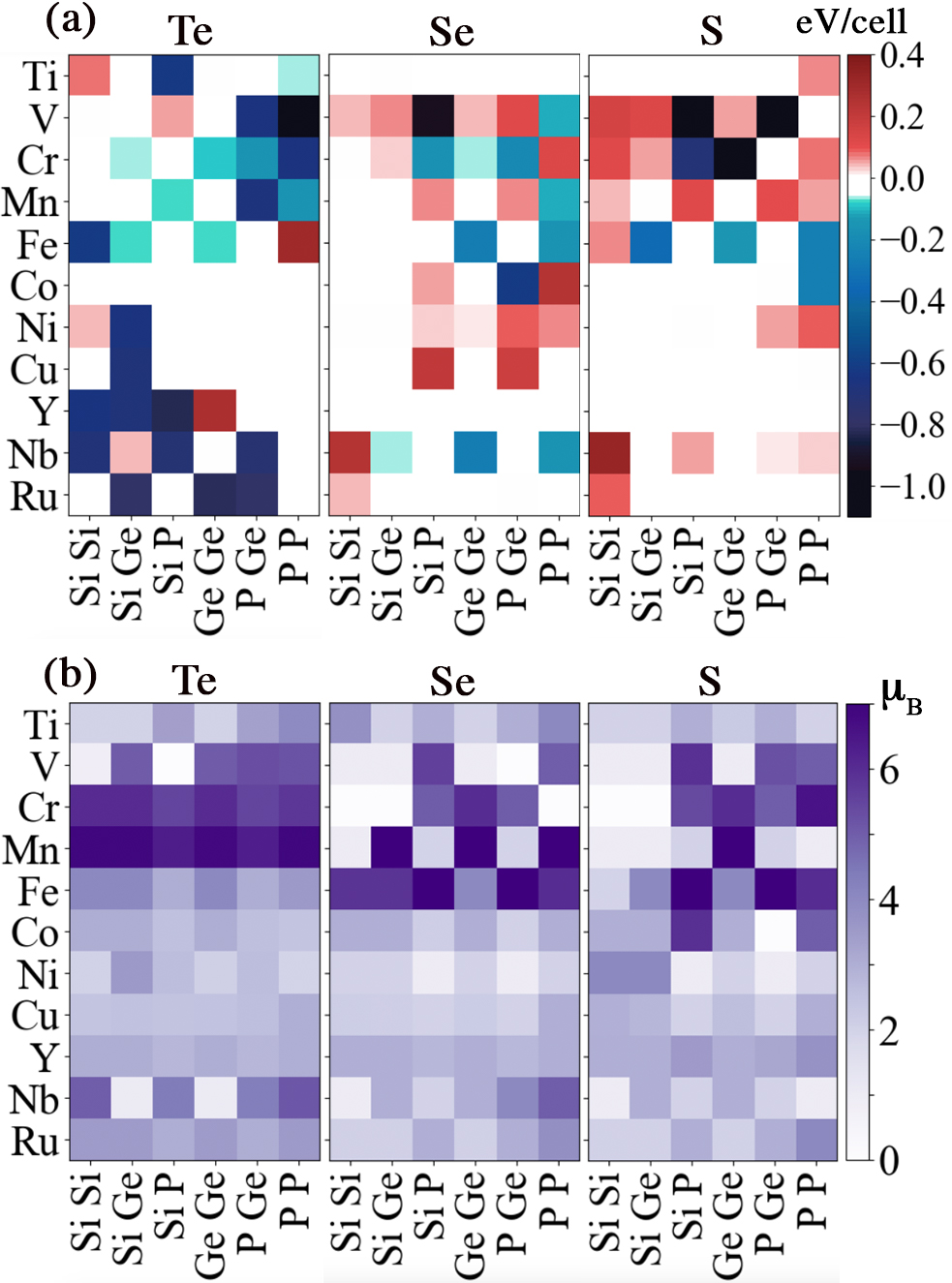}
  \caption{ \label{fig2} (a) Energy difference between parallel and anti-parallel spin configurations 
  ($E_{\text{parallel}}-E_{\text{anti-parallel}}$ in eV$/$unit cell) of A$_2$B$_2$X$_6$ structures. 
(b) Magnetic moment per unit cell (in $\mu_B$) for each A$_2$B$_2$X$_6$ structure at the lowest energy spin configuration. 
The occupation of the two B sites is shown on the horizontal axis while that of one of the A site is shown on the vertical axis.
}
\end{figure}

We find that the non-spin-polarized configuration has the highest energy for all the structures considered.
That is, all structures prefer either parallel or anti-parallel ordering in the A plane. 
%
Fig.~\ref{fig2}(a) shows the energy difference of parallel and anti-parallel spin configurations. 
Negative (positive) energy difference means the parallel (anti-parallel) is more stable.
We note that, because of the supercell size limit, we do not consider more complex spin configurations in this study.
For example, the lowest-energy spin configuration of Cr$_2$Si$_2$Te$_6$ was reported to be zigzag anti-ferromagnetic type~\cite{sivadas2015}.
%
%
%
Total magnetic moments for the lowest energy spin configuration of each structure are presented in Fig.~\ref{fig2}(b). 
We find that only atoms in the A sites show finite magnetic moments, while the moments in the B and X sites are small. 
Distinct patterns for regions of high and low magnetic moments are observed for X = Te, Se and S in Fig.~\ref{fig2}(b). 
Structures created by substituting non-magnetic atoms at the A site, such as Cu, have small variations in their relatively small magnetic moments, as seen in the rows of Fig.~\ref{fig2}(b).
However, substitutions of magnetic atoms, such as Mn, result in a set of structures with a large variation in the magnetic moment, with a much larger upper limit to the range of values observed.

Both the magnetic order and magnetic moment are sensitive to the occupancy of B and X sites, 
even though the atoms in these sites have negligible contribution to the overall magnetic moment. 
Atoms in the X sites strongly mediate the magnetic coupling between neighboring A sites~\cite{sivadas2015}. 
Atoms at the B sites can affect the relative positions of A and X sites.
Direct exchange between first nearest neighbor A sites competes with super-exchange interactions mediated by the $\textit{p}$-orbitals at the X sites.  
The ground state magnetic order is determined by the interplay between first, second and third nearest neighbor interactions.
Changing the identity of one of the A, B or X sites affects the interplay between the direct exchange and super-exchange interactions.  
Recent work has shown that applying strain to the Cr$_2$Si$_2$Te$_6$ lattice 
tunes the first nearest neighbor interaction, resulting in a change in the magnetic ground state
 from zig-zag antiferromagnetic to ferromagnetic~\cite{sivadas2015}. 
Our work demonstrates that tuning the composition of the A$_2$B$_2$X$_6$ lattice can have an equivalent effect. 
For instance, whereas X=Te structures show more parallel ($\bar{\bar{P}}$) than anti-parallel (anti-$\bar{\bar{P}}$) spin-configurations with lower energy,
there is a clear change when X $=$ Se or S.  
As X moves up the periodic table, there are increasingly more regions of anti-parallel spin configuration, 
as well as regions in which $\bar{\bar{P}}$ and anti-$\bar{\bar{P}}$ are degenerate.
In particular, we find that the distance between nearest neighbor A and X sites, 
as well as two adjacent X sites is linked to the magnitude of the magnetic moment (see Supplemental Materials for details).


%
%

\begin{figure}[h!]
\includegraphics[width=0.477\textwidth, keepaspectratio]{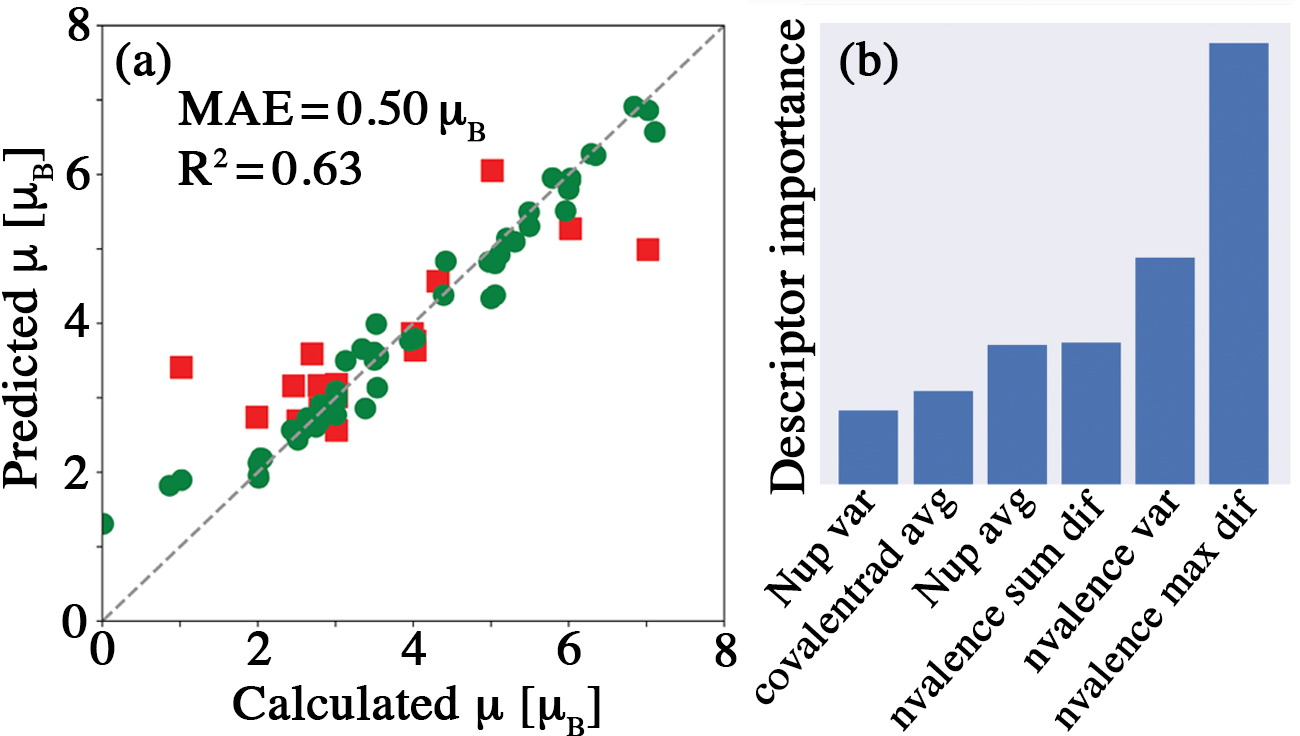}
\caption{\label{fig3} ML predictions of magnetic moments of A$_2$B$_2$X$_6$ structures. 
(a) Extra trees model performance for the magnetic moment (in $\mu_B$) prediction. A subset of structures for X = Te are displayed. 
The red squares indicate the test data, the green circles show the training data.
(b) Top six descriptors for the extra trees prediction of the magnetic moment. 
The size of the bar indicates relative descriptor importance (see text for details).}
\end{figure}

We use extra trees regression~\cite{hastie} to approximate 
the relationship between the total magnetic moment and a set of descriptors designed for magnetic property prediction 
(see Supplemental Materials). 
Training and test data are considered for the X = Te, Se, and S structures individually.  
The model performance for X = Te is shown in Fig.~\ref{fig3}(a). 
We find reasonable prediction performance for X = Te that deteriorates for X = Se and is even worse for X = S.
This suggests that our model, along with the set of descriptors used to predict X = Te structures, does not generalize well.  
This could arise due to the fact that there are more structures that have degenerate $\bar{\bar{P}}$ and anti-$\bar{\bar{P}}$ spin configurations if X=Se and S than for X = Te.
Nevertheless, subgroup discovery can be exploited to learn more about these systems~\cite{scheffler2017}, implying
that the identity of the X site strongly affects the magnetic properties of the structures.

%
%

%

\begin{figure}[h!]
  \includegraphics[width=0.48\textwidth, keepaspectratio]{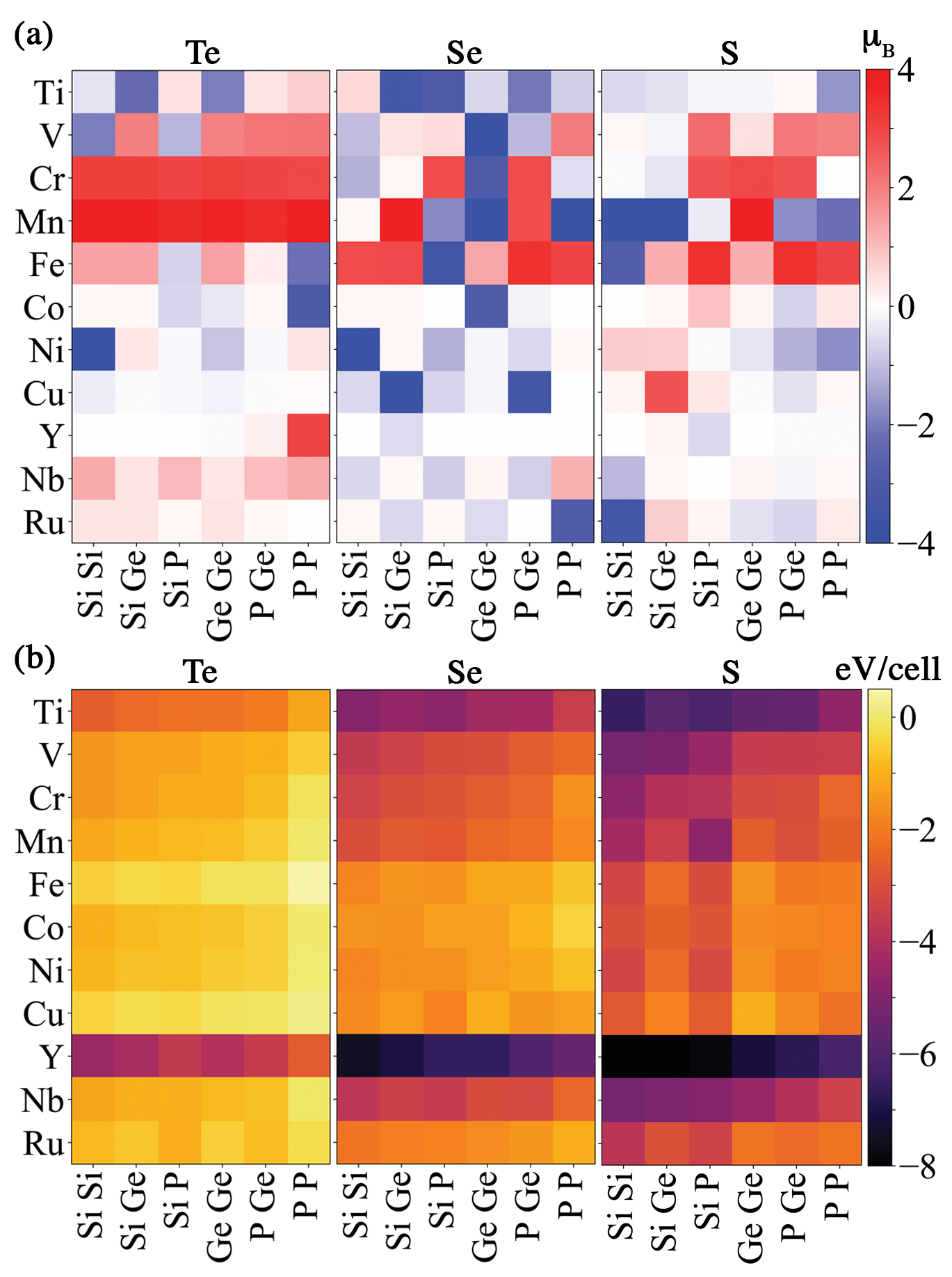}
  \caption{\label{fig4} (a) Local magnetic moment of the transition metal A site, A$_{\text{TM}}$ (in $\mu_{\text{B}}$). 
  (b) Formation energy (in eV/cell) for A$_2$B$_2$X$_6$ structures at the lowest energy spin configuration. 
  Conventions are the same as in Fig.~\ref{fig2}.
   }
\end{figure}

Determining which descriptors are most important for making good predictions of a property
can be exploited for knowledge discovery, especially when a large number of descriptors are available 
but their relationships with the target property are not known~\cite{reshef}.   
Fig.~\ref{fig3}(b) shows the descriptor importances~\cite{scikitlearn} as derived from extra trees regression. 
It shows that the `\emph{the number of valence electrons}' [``nvalence max dif'' in Fig.~\ref{fig3}(b)], 
`\emph{the average covalent radius}' [``covalentrad avg'' in Fig.~\ref{fig3}(b)]
and the `\emph{average number of spin up electrons}' [``Nup avg'' in Fig.~\ref{fig3}(b)],
linked to the atomic dipole magnetic moment,
are among the top six descriptors in the set examined.  
The magnetic moment per unit cell is a function of the magnetic moments of the individual atoms in the unit cell.
We examine the local magnetic moments at the A sites to determine how the magnetic moment per unit cell is constructed.
%
%
%
The local magnetic moment at the A sites (A$_{\text{Cr}}$ and A$_{\text{TM}}$) can be different from the atomic dipole magnetic moment of the corresponding element.  
For instance, while the atomic magnetic moment of Cr$^{3+}$ is 3~$\mu_B$, the local magnetic moment at  A$_{\text{Cr}}$ fluctuates from 2.7 to 3.2~$\mu_B$. 
Fig.~\ref{fig4} (a) shows the local magnetic moment at A$_{\text{TM}}$.
%
%

\subsection{Formation energy}

In addition to identifying structures with specific magnetic properties, 
the ability to screen for chemical stability is also important. 
DFT-calculated formation energies (for the lowest energy spin configuration) are shown in Fig.~\ref{fig4} (b).  
Structures comprising certain elements, such as Y, decrease the formation energy considerably in comparison to those without it.  
Certain transition metals, such as Cu, tend to destabilize the (CrA)B$_2$X$_6$ structures. 
The formation energy becomes more negative as the substituted atom at the A site goes from the
left to the right of the first and second row of transition metal elements in the Periodic Table.
This is linked to the filling of the \textit{d}-orbital, where elements with a filled \textit{d}-orbital do not form chemical bonds
with other elements.
Varying the composition at the B site does not appear to have a strong impact on the formation energy (see Supplemental Materials, Fig. S1).
Changing the X site from Te to Se and then S results in the overall trend of decreasing formation energy. 
%
%
%
%
%


To exploit the trends in the formation energy data, we use statistical models to predict the formation energy and 
to infer structure-property relationships. 
%
We find that some descriptors, such as the atomic dipole polarizability, are strongly correlated with the formation energy,
and are therefore 
important in generating good ML predictions.
Since useful descriptors are not always revealed in an analysis of the Pearson correlation coefficient~\cite{reshef},
we consider other methods to learn descriptor importances such as the extra trees model~\cite{scikitlearn}.
 \begin{figure}[h!]
  \includegraphics[width=0.48\textwidth, keepaspectratio]{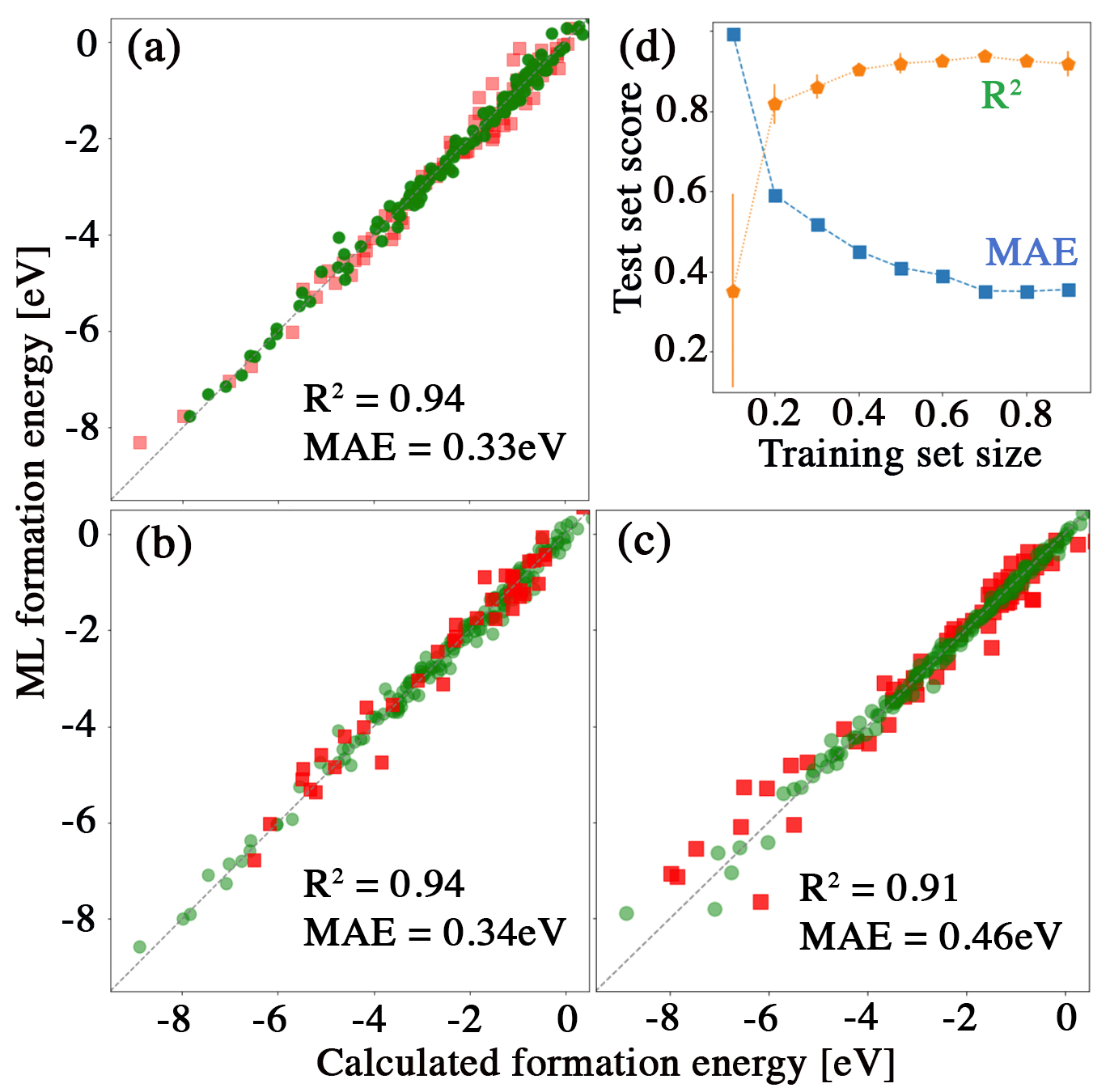}
  \caption{\label{fig5}Formation energy prediction performance of (a) kernel ridge regression, 
         (b) deep neural network regression and (c) extra trees regression. Red squares are
         test data and green circles training data.
         (d) Performance of the extra trees regression model on the test data as the training set size increases, 
         in terms of the R$^2$ and mean absolute error (MAE) scores.
 }
\end{figure}
Using the ML models to predict the formation energy of A$_2$B$_2$X$_6$ structures  
permits the quick calculation of the formation energy for a large set of compounds. 
Whereas DFT calculations of 10$^4$ structures
could take up to 1 million CPU hours, the ML prediction takes a few seconds. 
Fig.~\ref{fig5}(a) shows the prediction performance for kernel ridge regression using a gaussian kernel. 
Fig.~\ref{fig5}(b) shows the performance of a neural network~\footnote{The deep neural network used in this study is implemented by tensorflow. It is comprised of 3 hidden layers with sizes 10, 30 and 10 units} while
Fig.~\ref{fig5}(c) shows the performance of the extra forests regression. 
Both training set and test set results are displayed, as well as the 
test scores for kernel ridge regression, extra trees regression, and neural network regression.

\begin{figure*}
\centering
{\includegraphics[width=6.90in]{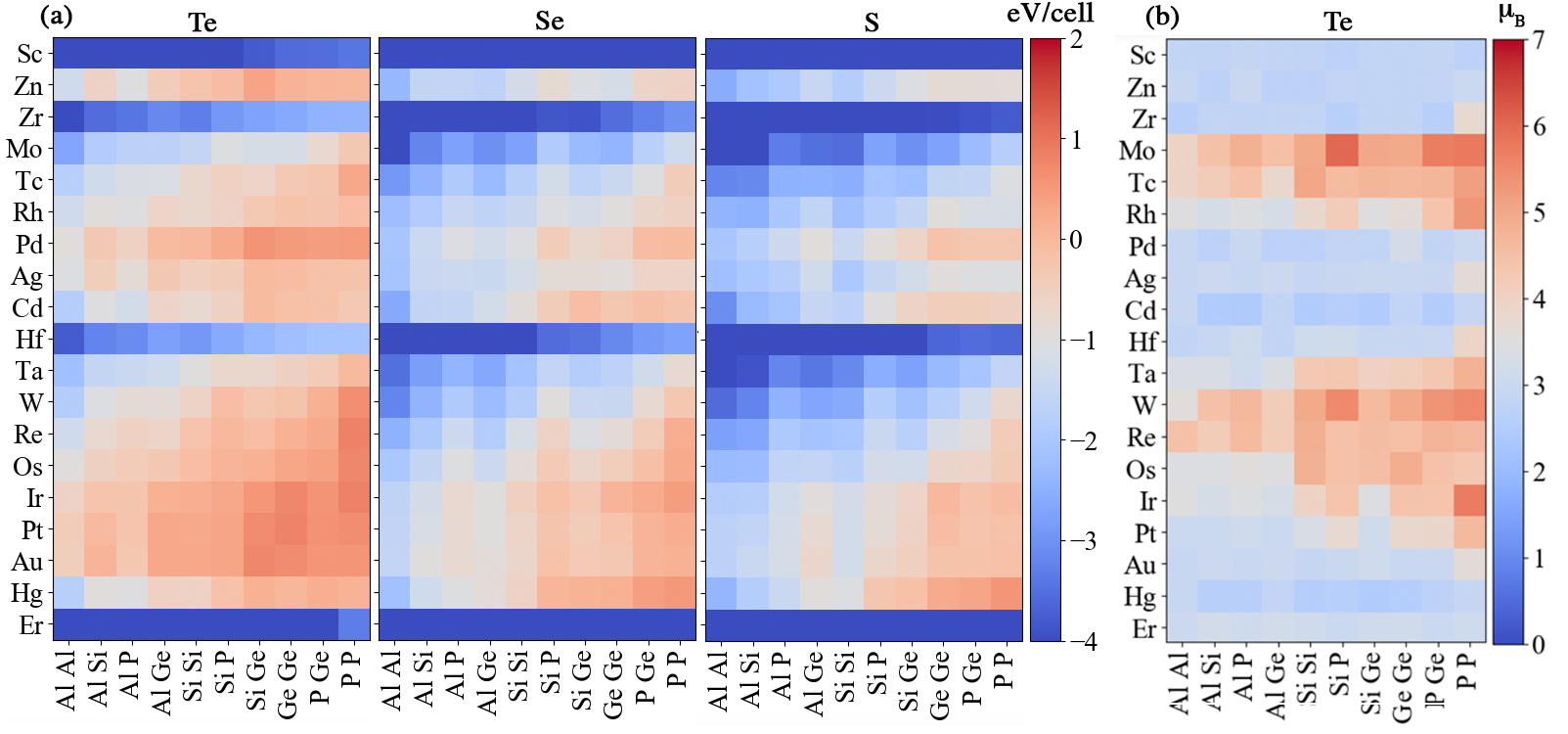}}
  \caption{\label{fig6} (a) ML predicted formation energies (in eV/cell) for a wide range of substitutions that      
   were not included in the DFT data set covering 4,223 new structures (570 are shown here). 
  (b) ML predicted magnetic moments (in $\mu_B$) for a wide range of substitutions that      
   were not included in the DFT data set covering 4,223 new structures (190 are shown here for X=Te). 
   Conventions same as in Fig.~\ref{fig2}.
    }
\end{figure*}

Further analysis (see Supplemental Materials) 
shows that the `\emph{variance in the ionization energy of atoms}' and the `\emph{average number of valence electrons}' are the two most important descriptors in the set examined. 
This demonstrates a  link between the formation energy and the atomic ionization energy, 
emanating from the increased atomic ionizability which produces stronger chemical bonding.
In addition, the number of valence electrons is linked to the number of electrons available for bonding. 
For instance, substitutions by atoms with a filled outer orbital shell
will create less stable bonds, leading to chemical instability.
%
The ability of our models to generalize is demonstrated by the high scores on the test data. 
We further examined how the test set performance varies with the training set size.  
Fig.~\ref{fig5}(d)
 shows test scores as a function of training set size using extra trees regression. 
The test score reaches a plateau at about a training set size of 40\%, with test score (R$^2$) as high as 0.91.
%

\subsection{High-throughput screening using ML models}
We can use our trained ML models to make predictions on a wide range of structures not included in the 
original DFT data set. 
Thus far, we have used our ML models to estimate the formation energy for an additional 4,223 A$_2$B$_2$X$_6$ structures, constructed as follows:
(i) For A site substitutions, we considered transition metals not used in the DFT dataset. 
(ii) We included Al, Sn and Pb in the set of atomic substitutions for B sites (not shown). 
(iii) For the X sites, we added O to our previous choice of S, Se and Te.
The resulting predictions, partly shown in Fig.~\ref{fig6}(a),  
provide a means to quickly screen a large data set of structures for chemical stability. 
For instance, our ML predictions suggest that structures based on Er, Ta, Hf, Mo, Zr, and Sc in the A site and Al in the B site are likely to be stable and thus good candidates for further exploration.

Magnetic moment predictions are shown in Fig.~\ref{fig6}(b).  
From the results of the ML predictions we select structures with formation energies below -1 eV and magnetic moments above 5~$\mu_B$ (for X=Te only). From the 4,223 predictions, we obtained 40 that satisfied our constraints. 15 of these were randomly selected for verification with DFT. 
5 of these 15 structures were confirmed to have the expected properties within uncertainty.  
These 15 structures were then added to the training data to build an improved model for predicting magnetic moment.  
A second iteration of prediction and verification by DFT generated three structures, all of which satisfied the constraints within uncertainty: (CrTc)(SiSn)Te$_6$, (CrTc)Sn$_2$Te$_6$, Cr$_2$(SiP)Te$_6$.


\section{Conclusion}
We presented evidence that the magnetic properties of A$_2$B$_2$X$_6$ monolayer structures can be tuned by making atomic substitutions at A, B, and X sites.  
This provides a novel framework for investigating the microscopic origin of magnetic order of 2D layered materials 
and could lead to insights into magnetism in systems of reduced dimension~\cite{huang2017, cava2017}. 
Our work represents a path toward tailoring magnetic properties of materials for applications in spintronics and data storage~\cite{han2016}.  
We showed that ML methods are promising tools for predicting the magnetic properties of 2D magnetic materials. 
In particular, our data-driven approach highlights the importance of the X site in determining the magnetic order
of the structure. 
Changing the composition of the A$_2$B$_2$X$_6$ structure alters the inter-atomic distances and the identity of electronic orbitals. 
This impacts the interplay between first, second and third nearest neighbor exchange interactions,
which determines the magnetic order.

One goal of this work was to find magnetic 2D materials that are also thermodynamically stable.
ML models were trained to predict chemical stability that allow the rapid screening of a large number of possible structures.
We showed that the chemical stability of A$_2$B$_2$X$_6$ structures based on Cr$_2$Ge$_2$Te$_6$ can be tuned by making atomic substitutions.   
Examples of structures that satisfy both magnetic moment and formation energy requirements include the following:
(CrTc)(SiSn)Te$_6$ and (CrTc)Sn$_2$Te$_6$, not included in our original DFT database. In addition, we found structures in our set of DFT calculations that also satisfied our requirements: Cr$_2$(SiP)Te$_6$, (TiCr)(SiP)Te$_6$, (YCr)Ge$_2$S$_6$ and (NbCr)Si$_2$Te$_6$.

This work provides the impetus for further exploration of structures with other architectures not considered here, that is, with more complex atomic substitutions beyond 1 in 2 replacement of Cr atoms at the A site.
We estimate a total number of at least 3$\times$10$^4$ structures of the A$_2$B$_2$X$_6$ type described in Fig.~\ref{fig1}.
A computationally efficient estimation of the magnetic properties and formation energy is required to quickly explore this vast chemical space. 
We also expect the ML methods explored here, with proper modification,
 to allow an efficient exploration of other families of 2D magnets, such as
CrI$_3$, CrOCl and Fe$_3$GeTe$_2$~\cite{chen2013, huang2017, sun2018}. 

  

\begin{acknowledgments}
We thank Marios Mattheakis, Daniel Larson, Robert Hoyt, Matthew Montemore, Sadas Shankar, Ekin Dogus Cubuk, Pavlos Protopapas and Vinothan Manoharan for helpful discussions.
For the calculations we used the Extreme Science and Engineering Discovery Environment (XSEDE), which is supported by National Science Foundation (grant number ACI-1548562)
and the Odyssey cluster supported by the FAS Division of Science, Research Computing Group at Harvard University.
T.D.R. is supported by the Harvard Future Faculty Leaders Postdoctoral Fellowship.
We acknowledge support from ARO MURI Award W911NF-14-0247. 
\end{acknowledgments}



\bibliographystyle{apsrev4-1}


\begin{thebibliography}{48}%
\makeatletter
\providecommand \@ifxundefined [1]{%
 \@ifx{#1\undefined}
}%
\providecommand \@ifnum [1]{%
 \ifnum #1\expandafter \@firstoftwo
 \else \expandafter \@secondoftwo
 \fi
}%
\providecommand \@ifx [1]{%
 \ifx #1\expandafter \@firstoftwo
 \else \expandafter \@secondoftwo
 \fi
}%
\providecommand \natexlab [1]{#1}%
\providecommand \enquote  [1]{``#1''}%
\providecommand \bibnamefont  [1]{#1}%
\providecommand \bibfnamefont [1]{#1}%
\providecommand \citenamefont [1]{#1}%
\providecommand \href@noop [0]{\@secondoftwo}%
\providecommand \href [0]{\begingroup \@sanitize@url \@href}%
\providecommand \@href[1]{\@@startlink{#1}\@@href}%
\providecommand \@@href[1]{\endgroup#1\@@endlink}%
\providecommand \@sanitize@url [0]{\catcode `\\12\catcode `\$12\catcode
  `\&12\catcode `\#12\catcode `\^12\catcode `\_12\catcode `\%12\relax}%
\providecommand \@@startlink[1]{}%
\providecommand \@@endlink[0]{}%
\providecommand \url  [0]{\begingroup\@sanitize@url \@url }%
\providecommand \@url [1]{\endgroup\@href {#1}{\urlprefix }}%
\providecommand \urlprefix  [0]{URL }%
\providecommand \Eprint [0]{\href }%
\providecommand \doibase [0]{http://dx.doi.org/}%
\providecommand \selectlanguage [0]{\@gobble}%
\providecommand \bibinfo  [0]{\@secondoftwo}%
\providecommand \bibfield  [0]{\@secondoftwo}%
\providecommand \translation [1]{[#1]}%
\providecommand \BibitemOpen [0]{}%
\providecommand \bibitemStop [0]{}%
\providecommand \bibitemNoStop [0]{.\EOS\space}%
\providecommand \EOS [0]{\spacefactor3000\relax}%
\providecommand \BibitemShut  [1]{\csname bibitem#1\endcsname}%
\let\auto@bib@innerbib\@empty
\bibitem [{\citenamefont {Bhimanapati}\ \emph {et~al.}(2015)\citenamefont
  {Bhimanapati}, \citenamefont {Lin}, \citenamefont {Meunier}, \citenamefont
  {Jung}, \citenamefont {Cha}, \citenamefont {Das}, \citenamefont {Xiao},
  \citenamefont {Son}, \citenamefont {Strano}, \citenamefont {Cooper} \emph
  {et~al.}}]{robinson2015}%
  \BibitemOpen
  \bibfield  {author} {\bibinfo {author} {\bibfnamefont {G.~R.}\ \bibnamefont
  {Bhimanapati}}, \bibinfo {author} {\bibfnamefont {Z.}~\bibnamefont {Lin}},
  \bibinfo {author} {\bibfnamefont {V.}~\bibnamefont {Meunier}}, \bibinfo
  {author} {\bibfnamefont {Y.}~\bibnamefont {Jung}}, \bibinfo {author}
  {\bibfnamefont {J.}~\bibnamefont {Cha}}, \bibinfo {author} {\bibfnamefont
  {S.}~\bibnamefont {Das}}, \bibinfo {author} {\bibfnamefont {D.}~\bibnamefont
  {Xiao}}, \bibinfo {author} {\bibfnamefont {Y.}~\bibnamefont {Son}}, \bibinfo
  {author} {\bibfnamefont {M.~S.}\ \bibnamefont {Strano}}, \bibinfo {author}
  {\bibfnamefont {V.~R.}\ \bibnamefont {Cooper}},  \emph {et~al.},\ }\href@noop
  {} {\bibfield  {journal} {\bibinfo  {journal} {ACS Nano}\ }\textbf {\bibinfo
  {volume} {9}},\ \bibinfo {pages} {11509} (\bibinfo {year}
  {2015})}\BibitemShut {NoStop}%
\bibitem [{\citenamefont {Cheon}\ \emph {et~al.}(2017)\citenamefont {Cheon},
  \citenamefont {Duerloo}, \citenamefont {Sendek}, \citenamefont {Porter},
  \citenamefont {Chen},\ and\ \citenamefont {Reed}}]{cheon2017}%
  \BibitemOpen
  \bibfield  {author} {\bibinfo {author} {\bibfnamefont {G.}~\bibnamefont
  {Cheon}}, \bibinfo {author} {\bibfnamefont {K.-A.~N.}\ \bibnamefont
  {Duerloo}}, \bibinfo {author} {\bibfnamefont {A.~D.}\ \bibnamefont {Sendek}},
  \bibinfo {author} {\bibfnamefont {C.}~\bibnamefont {Porter}}, \bibinfo
  {author} {\bibfnamefont {Y.}~\bibnamefont {Chen}}, \ and\ \bibinfo {author}
  {\bibfnamefont {E.~J.}\ \bibnamefont {Reed}},\ }\href@noop {} {\bibfield
  {journal} {\bibinfo  {journal} {Nano Lett.}\ }\textbf {\bibinfo {volume}
  {17}},\ \bibinfo {pages} {1915} (\bibinfo {year} {2017})}\BibitemShut
  {NoStop}%
\bibitem [{\citenamefont {Mounet}\ \emph {et~al.}(2018)\citenamefont {Mounet},
  \citenamefont {Gibertini}, \citenamefont {Schwaller}, \citenamefont {Campi},
  \citenamefont {Merkys}, \citenamefont {Marrazzo}, \citenamefont {Sohier},
  \citenamefont {Castelli}, \citenamefont {Cepellotti}, \citenamefont {Pizzi}
  \emph {et~al.}}]{Marzari}%
  \BibitemOpen
  \bibfield  {author} {\bibinfo {author} {\bibfnamefont {N.}~\bibnamefont
  {Mounet}}, \bibinfo {author} {\bibfnamefont {M.}~\bibnamefont {Gibertini}},
  \bibinfo {author} {\bibfnamefont {P.}~\bibnamefont {Schwaller}}, \bibinfo
  {author} {\bibfnamefont {D.}~\bibnamefont {Campi}}, \bibinfo {author}
  {\bibfnamefont {A.}~\bibnamefont {Merkys}}, \bibinfo {author} {\bibfnamefont
  {A.}~\bibnamefont {Marrazzo}}, \bibinfo {author} {\bibfnamefont
  {T.}~\bibnamefont {Sohier}}, \bibinfo {author} {\bibfnamefont {I.~E.}\
  \bibnamefont {Castelli}}, \bibinfo {author} {\bibfnamefont {A.}~\bibnamefont
  {Cepellotti}}, \bibinfo {author} {\bibfnamefont {G.}~\bibnamefont {Pizzi}},
  \emph {et~al.},\ }\href@noop {} {\bibfield  {journal} {\bibinfo  {journal}
  {Nat. Nanotechnology}\ }\textbf {\bibinfo {volume} {13}},\ \bibinfo {pages}
  {246} (\bibinfo {year} {2018})}\BibitemShut {NoStop}%
\bibitem [{\citenamefont {Chen}\ \emph
  {et~al.}(2013{\natexlab{a}})\citenamefont {Chen}, \citenamefont {Santos},
  \citenamefont {Zhu}, \citenamefont {Kaxiras},\ and\ \citenamefont
  {Zhang}}]{chen2013tuning}%
  \BibitemOpen
  \bibfield  {author} {\bibinfo {author} {\bibfnamefont {W.}~\bibnamefont
  {Chen}}, \bibinfo {author} {\bibfnamefont {E.}~\bibnamefont {Santos}},
  \bibinfo {author} {\bibfnamefont {W.}~\bibnamefont {Zhu}}, \bibinfo {author}
  {\bibfnamefont {E.}~\bibnamefont {Kaxiras}}, \ and\ \bibinfo {author}
  {\bibfnamefont {Z.}~\bibnamefont {Zhang}},\ }\href@noop {} {\bibfield
  {journal} {\bibinfo  {journal} {Nano Lett.}\ }\textbf {\bibinfo {volume}
  {13}},\ \bibinfo {pages} {509} (\bibinfo {year}
  {2013}{\natexlab{a}})}\BibitemShut {NoStop}%
\bibitem [{\citenamefont {Nourbakhsh}\ \emph {et~al.}(2016)\citenamefont
  {Nourbakhsh}, \citenamefont {Zubair}, \citenamefont {Sajjad}, \citenamefont
  {Tavakkoli~KG}, \citenamefont {Chen}, \citenamefont {Fang}, \citenamefont
  {Ling}, \citenamefont {Kong}, \citenamefont {Dresselhaus}, \citenamefont
  {Kaxiras}, \citenamefont {Berggren}, \citenamefont {Antoniadia},\ and\
  \citenamefont {T.}}]{nourbakhsh2016mos2}%
  \BibitemOpen
  \bibfield  {author} {\bibinfo {author} {\bibfnamefont {A.}~\bibnamefont
  {Nourbakhsh}}, \bibinfo {author} {\bibfnamefont {A.}~\bibnamefont {Zubair}},
  \bibinfo {author} {\bibfnamefont {R.}~\bibnamefont {Sajjad}}, \bibinfo
  {author} {\bibfnamefont {A.}~\bibnamefont {Tavakkoli~KG}}, \bibinfo {author}
  {\bibfnamefont {W.}~\bibnamefont {Chen}}, \bibinfo {author} {\bibfnamefont
  {S.}~\bibnamefont {Fang}}, \bibinfo {author} {\bibfnamefont {X.}~\bibnamefont
  {Ling}}, \bibinfo {author} {\bibfnamefont {J.}~\bibnamefont {Kong}}, \bibinfo
  {author} {\bibfnamefont {M.}~\bibnamefont {Dresselhaus}}, \bibinfo {author}
  {\bibfnamefont {E.}~\bibnamefont {Kaxiras}}, \bibinfo {author} {\bibfnamefont
  {K.}~\bibnamefont {Berggren}}, \bibinfo {author} {\bibfnamefont
  {D.}~\bibnamefont {Antoniadia}}, \ and\ \bibinfo {author} {\bibfnamefont
  {P.}~\bibnamefont {T.}},\ }\href@noop {} {\bibfield  {journal} {\bibinfo
  {journal} {Nano Lett.}\ }\textbf {\bibinfo {volume} {16}},\ \bibinfo {pages}
  {7798} (\bibinfo {year} {2016})}\BibitemShut {NoStop}%
\bibitem [{\citenamefont {Novoselov}\ \emph {et~al.}(2016)\citenamefont
  {Novoselov}, \citenamefont {Mishchenko}, \citenamefont {Carvalho},\ and\
  \citenamefont {Neto}}]{novoselov20162d}%
  \BibitemOpen
  \bibfield  {author} {\bibinfo {author} {\bibfnamefont {K.}~\bibnamefont
  {Novoselov}}, \bibinfo {author} {\bibfnamefont {A.}~\bibnamefont
  {Mishchenko}}, \bibinfo {author} {\bibfnamefont {A.}~\bibnamefont
  {Carvalho}}, \ and\ \bibinfo {author} {\bibfnamefont {A.~C.}\ \bibnamefont
  {Neto}},\ }\href@noop {} {\bibfield  {journal} {\bibinfo  {journal}
  {Science}\ }\textbf {\bibinfo {volume} {353}},\ \bibinfo {pages} {aac9439}
  (\bibinfo {year} {2016})}\BibitemShut {NoStop}%
\bibitem [{\citenamefont {Zeng}\ \emph {et~al.}(2016)\citenamefont {Zeng},
  \citenamefont {Chen}, \citenamefont {Cui}, \citenamefont {Zhang},\ and\
  \citenamefont {Zhang}}]{zeng2016enhanced}%
  \BibitemOpen
  \bibfield  {author} {\bibinfo {author} {\bibfnamefont {J.}~\bibnamefont
  {Zeng}}, \bibinfo {author} {\bibfnamefont {W.}~\bibnamefont {Chen}}, \bibinfo
  {author} {\bibfnamefont {P.}~\bibnamefont {Cui}}, \bibinfo {author}
  {\bibfnamefont {D.}~\bibnamefont {Zhang}}, \ and\ \bibinfo {author}
  {\bibfnamefont {Z.}~\bibnamefont {Zhang}},\ }\href@noop {} {\bibfield
  {journal} {\bibinfo  {journal} {Phys. Rev. B}\ }\textbf {\bibinfo {volume}
  {94}},\ \bibinfo {pages} {235425} (\bibinfo {year} {2016})}\BibitemShut
  {NoStop}%
\bibitem [{\citenamefont {Choi}\ \emph {et~al.}(2017)\citenamefont {Choi},
  \citenamefont {Cui}, \citenamefont {Chen}, \citenamefont {Cho},\ and\
  \citenamefont {Zhang}}]{choi2017atomistic}%
  \BibitemOpen
  \bibfield  {author} {\bibinfo {author} {\bibfnamefont {J.}~\bibnamefont
  {Choi}}, \bibinfo {author} {\bibfnamefont {P.}~\bibnamefont {Cui}}, \bibinfo
  {author} {\bibfnamefont {W.}~\bibnamefont {Chen}}, \bibinfo {author}
  {\bibfnamefont {J.}~\bibnamefont {Cho}}, \ and\ \bibinfo {author}
  {\bibfnamefont {Z.}~\bibnamefont {Zhang}},\ }\href@noop {} {\bibfield
  {journal} {\bibinfo  {journal} {Wiley Interdiscip. Rev. Comput. Mol. Sci.}\
  }\textbf {\bibinfo {volume} {7}} (\bibinfo {year} {2017})}\BibitemShut
  {NoStop}%
\bibitem [{\citenamefont {Chen}\ \emph {et~al.}(2017)\citenamefont {Chen},
  \citenamefont {Yang}, \citenamefont {Zhang},\ and\ \citenamefont
  {Kaxiras}}]{chen2017properties}%
  \BibitemOpen
  \bibfield  {author} {\bibinfo {author} {\bibfnamefont {W.}~\bibnamefont
  {Chen}}, \bibinfo {author} {\bibfnamefont {Y.}~\bibnamefont {Yang}}, \bibinfo
  {author} {\bibfnamefont {Z.}~\bibnamefont {Zhang}}, \ and\ \bibinfo {author}
  {\bibfnamefont {E.}~\bibnamefont {Kaxiras}},\ }\href@noop {} {\bibfield
  {journal} {\bibinfo  {journal} {2D Mater.}\ }\textbf {\bibinfo {volume}
  {4}},\ \bibinfo {pages} {045001} (\bibinfo {year} {2017})}\BibitemShut
  {NoStop}%
\bibitem [{\citenamefont {Huo}\ and\ \citenamefont {Rupp}(2017)}]{mbtr}%
  \BibitemOpen
  \bibfield  {author} {\bibinfo {author} {\bibfnamefont {H.}~\bibnamefont
  {Huo}}\ and\ \bibinfo {author} {\bibfnamefont {M.}~\bibnamefont {Rupp}},\
  }\href@noop {} {\bibfield  {journal} {\bibinfo  {journal} {arXiv:1704.06439}\
  } (\bibinfo {year} {2017})}\BibitemShut {NoStop}%
\bibitem [{\citenamefont {Gilmer}\ \emph {et~al.}(2017)\citenamefont {Gilmer},
  \citenamefont {Schoenholz}, \citenamefont {Riley}, \citenamefont {Vinyals},\
  and\ \citenamefont {Dahl}}]{riley2017}%
  \BibitemOpen
  \bibfield  {author} {\bibinfo {author} {\bibfnamefont {J.}~\bibnamefont
  {Gilmer}}, \bibinfo {author} {\bibfnamefont {S.~S.}\ \bibnamefont
  {Schoenholz}}, \bibinfo {author} {\bibfnamefont {P.~F.}\ \bibnamefont
  {Riley}}, \bibinfo {author} {\bibfnamefont {O.}~\bibnamefont {Vinyals}}, \
  and\ \bibinfo {author} {\bibfnamefont {G.~E.}\ \bibnamefont {Dahl}},\ }in\
  \href {http://proceedings.mlr.press/v70/gilmer17a.html} {\emph {\bibinfo
  {booktitle} {Proceedings of the 34th International Conference on Machine
  Learning}}},\ \bibinfo {series} {Proceedings of Machine Learning Research},
  Vol.~\bibinfo {volume} {70},\ \bibinfo {editor} {edited by\ \bibinfo {editor}
  {\bibfnamefont {D.}~\bibnamefont {Precup}}\ and\ \bibinfo {editor}
  {\bibfnamefont {Y.~W.}\ \bibnamefont {Teh}}}\ (\bibinfo  {publisher} {PMLR},\
  \bibinfo {address} {International Convention Centre, Sydney, Australia},\
  \bibinfo {year} {2017})\ pp.\ \bibinfo {pages} {1263--1272}\BibitemShut
  {NoStop}%
\bibitem [{\citenamefont {Isayev}\ \emph {et~al.}(2017)\citenamefont {Isayev},
  \citenamefont {Oses}, \citenamefont {Toher}, \citenamefont {Gossett},
  \citenamefont {Curtarolo},\ and\ \citenamefont {Tropsha}}]{tropsha2017}%
  \BibitemOpen
  \bibfield  {author} {\bibinfo {author} {\bibfnamefont {O.}~\bibnamefont
  {Isayev}}, \bibinfo {author} {\bibfnamefont {C.}~\bibnamefont {Oses}},
  \bibinfo {author} {\bibfnamefont {C.}~\bibnamefont {Toher}}, \bibinfo
  {author} {\bibfnamefont {E.}~\bibnamefont {Gossett}}, \bibinfo {author}
  {\bibfnamefont {S.}~\bibnamefont {Curtarolo}}, \ and\ \bibinfo {author}
  {\bibfnamefont {A.}~\bibnamefont {Tropsha}},\ }\href@noop {} {\bibfield
  {journal} {\bibinfo  {journal} {Nat. Commun.}\ }\textbf {\bibinfo {volume}
  {8}},\ \bibinfo {pages} {15679} (\bibinfo {year} {2017})}\BibitemShut
  {NoStop}%
\bibitem [{\citenamefont {Huang}\ \emph {et~al.}(2017)\citenamefont {Huang},
  \citenamefont {Clark}, \citenamefont {Navarro-Moratalla}, \citenamefont
  {Klein}, \citenamefont {Cheng}, \citenamefont {Seyler}, \citenamefont
  {Zhong}, \citenamefont {Schmidgall}, \citenamefont {McGuire}, \citenamefont
  {Cobden}, \citenamefont {Yao}, \citenamefont {Xiao}, \citenamefont
  {Jarillo-Herrero},\ and\ \citenamefont {Xu}}]{huang2017}%
  \BibitemOpen
  \bibfield  {author} {\bibinfo {author} {\bibfnamefont {B.}~\bibnamefont
  {Huang}}, \bibinfo {author} {\bibfnamefont {G.}~\bibnamefont {Clark}},
  \bibinfo {author} {\bibfnamefont {E.}~\bibnamefont {Navarro-Moratalla}},
  \bibinfo {author} {\bibfnamefont {D.~R.}\ \bibnamefont {Klein}}, \bibinfo
  {author} {\bibfnamefont {R.}~\bibnamefont {Cheng}}, \bibinfo {author}
  {\bibfnamefont {K.~L.}\ \bibnamefont {Seyler}}, \bibinfo {author}
  {\bibfnamefont {D.}~\bibnamefont {Zhong}}, \bibinfo {author} {\bibfnamefont
  {E.}~\bibnamefont {Schmidgall}}, \bibinfo {author} {\bibfnamefont {M.~A.}\
  \bibnamefont {McGuire}}, \bibinfo {author} {\bibfnamefont {D.~H.}\
  \bibnamefont {Cobden}}, \bibinfo {author} {\bibfnamefont {W.}~\bibnamefont
  {Yao}}, \bibinfo {author} {\bibfnamefont {D.}~\bibnamefont {Xiao}}, \bibinfo
  {author} {\bibfnamefont {P.}~\bibnamefont {Jarillo-Herrero}}, \ and\ \bibinfo
  {author} {\bibfnamefont {X.}~\bibnamefont {Xu}},\ }\href@noop {} {\bibfield
  {journal} {\bibinfo  {journal} {Nature}\ }\textbf {\bibinfo {volume} {546}},\
  \bibinfo {pages} {270} (\bibinfo {year} {2017})}\BibitemShut {NoStop}%
\bibitem [{\citenamefont {Gong}\ \emph {et~al.}(2017)\citenamefont {Gong},
  \citenamefont {Li}, \citenamefont {Li}, \citenamefont {Ji}, \citenamefont
  {Stern}, \citenamefont {Xia}, \citenamefont {Cao}, \citenamefont {Bao},
  \citenamefont {Wang}, \citenamefont {Wang}, \citenamefont {Qiu},
  \citenamefont {Cava}, \citenamefont {Louie}, \citenamefont {Xia},\ and\
  \citenamefont {Zhang}}]{cava2017}%
  \BibitemOpen
  \bibfield  {author} {\bibinfo {author} {\bibfnamefont {C.}~\bibnamefont
  {Gong}}, \bibinfo {author} {\bibfnamefont {L.}~\bibnamefont {Li}}, \bibinfo
  {author} {\bibfnamefont {Z.}~\bibnamefont {Li}}, \bibinfo {author}
  {\bibfnamefont {H.}~\bibnamefont {Ji}}, \bibinfo {author} {\bibfnamefont
  {A.}~\bibnamefont {Stern}}, \bibinfo {author} {\bibfnamefont
  {Y.}~\bibnamefont {Xia}}, \bibinfo {author} {\bibfnamefont {T.}~\bibnamefont
  {Cao}}, \bibinfo {author} {\bibfnamefont {W.}~\bibnamefont {Bao}}, \bibinfo
  {author} {\bibfnamefont {C.}~\bibnamefont {Wang}}, \bibinfo {author}
  {\bibfnamefont {Y.}~\bibnamefont {Wang}}, \bibinfo {author} {\bibfnamefont
  {Z.~Q.}\ \bibnamefont {Qiu}}, \bibinfo {author} {\bibfnamefont {R.~J.}\
  \bibnamefont {Cava}}, \bibinfo {author} {\bibfnamefont {S.~G.}\ \bibnamefont
  {Louie}}, \bibinfo {author} {\bibfnamefont {J.}~\bibnamefont {Xia}}, \ and\
  \bibinfo {author} {\bibfnamefont {X.}~\bibnamefont {Zhang}},\ }\href@noop {}
  {\bibfield  {journal} {\bibinfo  {journal} {Nature}\ }\textbf {\bibinfo
  {volume} {546}},\ \bibinfo {pages} {265} (\bibinfo {year}
  {2017})}\BibitemShut {NoStop}%
\bibitem [{\citenamefont {Cui}\ \emph {et~al.}(2017)\citenamefont {Cui},
  \citenamefont {Choi}, \citenamefont {Chen}, \citenamefont {Zeng},
  \citenamefont {Shih}, \citenamefont {Li},\ and\ \citenamefont
  {Zhang}}]{cui2017contrasting}%
  \BibitemOpen
  \bibfield  {author} {\bibinfo {author} {\bibfnamefont {P.}~\bibnamefont
  {Cui}}, \bibinfo {author} {\bibfnamefont {J.}~\bibnamefont {Choi}}, \bibinfo
  {author} {\bibfnamefont {W.}~\bibnamefont {Chen}}, \bibinfo {author}
  {\bibfnamefont {J.}~\bibnamefont {Zeng}}, \bibinfo {author} {\bibfnamefont
  {C.}~\bibnamefont {Shih}}, \bibinfo {author} {\bibfnamefont {Z.}~\bibnamefont
  {Li}}, \ and\ \bibinfo {author} {\bibfnamefont {Z.}~\bibnamefont {Zhang}},\
  }\href@noop {} {\bibfield  {journal} {\bibinfo  {journal} {Nano Lett.}\
  }\textbf {\bibinfo {volume} {17}},\ \bibinfo {pages} {1097} (\bibinfo {year}
  {2017})}\BibitemShut {NoStop}%
\bibitem [{\citenamefont {Miyazato}\ \emph {et~al.}(2018)\citenamefont
  {Miyazato}, \citenamefont {Tanaka},\ and\ \citenamefont
  {Takahashi}}]{miyazato2018}%
  \BibitemOpen
  \bibfield  {author} {\bibinfo {author} {\bibfnamefont {I.}~\bibnamefont
  {Miyazato}}, \bibinfo {author} {\bibfnamefont {Y.}~\bibnamefont {Tanaka}}, \
  and\ \bibinfo {author} {\bibfnamefont {K.}~\bibnamefont {Takahashi}},\
  }\href@noop {} {\bibfield  {journal} {\bibinfo  {journal} {J. Phys. Condens.
  Matter}\ }\textbf {\bibinfo {volume} {30}},\ \bibinfo {pages} {06LT01}
  (\bibinfo {year} {2018})}\BibitemShut {NoStop}%
\bibitem [{\citenamefont {M{\"o}ller}\ \emph {et~al.}(2018)\citenamefont
  {M{\"o}ller}, \citenamefont {K{\"o}rner}, \citenamefont {Krugel},
  \citenamefont {Urban},\ and\ \citenamefont {Els{\"a}sser}}]{moller2018}%
  \BibitemOpen
  \bibfield  {author} {\bibinfo {author} {\bibfnamefont {J.~J.}\ \bibnamefont
  {M{\"o}ller}}, \bibinfo {author} {\bibfnamefont {W.}~\bibnamefont
  {K{\"o}rner}}, \bibinfo {author} {\bibfnamefont {G.}~\bibnamefont {Krugel}},
  \bibinfo {author} {\bibfnamefont {D.~F.}\ \bibnamefont {Urban}}, \ and\
  \bibinfo {author} {\bibfnamefont {C.}~\bibnamefont {Els{\"a}sser}},\ }\href
  {\doibase https://doi.org/10.1016/j.actamat.2018.03.051} {\bibfield
  {journal} {\bibinfo  {journal} {Acta Materialia}\ }\textbf {\bibinfo {volume}
  {153}},\ \bibinfo {pages} {53 } (\bibinfo {year} {2018})}\BibitemShut
  {NoStop}%
\bibitem [{\citenamefont {Miao}\ \emph {et~al.}(2018)\citenamefont {Miao},
  \citenamefont {Xu}, \citenamefont {Zhu}, \citenamefont {Zhou},\ and\
  \citenamefont {Sun}}]{sun2018}%
  \BibitemOpen
  \bibfield  {author} {\bibinfo {author} {\bibfnamefont {N.}~\bibnamefont
  {Miao}}, \bibinfo {author} {\bibfnamefont {B.}~\bibnamefont {Xu}}, \bibinfo
  {author} {\bibfnamefont {L.}~\bibnamefont {Zhu}}, \bibinfo {author}
  {\bibfnamefont {J.}~\bibnamefont {Zhou}}, \ and\ \bibinfo {author}
  {\bibfnamefont {Z.}~\bibnamefont {Sun}},\ }\href {\doibase
  10.1021/jacs.7b12976} {\bibfield  {journal} {\bibinfo  {journal} {Journal of
  the American Chemical Society}\ }\textbf {\bibinfo {volume} {140}},\ \bibinfo
  {pages} {2417} (\bibinfo {year} {2018})},\ \bibinfo {note} {pMID: 29400056},\
  \Eprint {http://arxiv.org/abs/https://doi.org/10.1021/jacs.7b12976}
  {https://doi.org/10.1021/jacs.7b12976} \BibitemShut {NoStop}%
\bibitem [{\citenamefont {Mermin}\ and\ \citenamefont
  {Wagner}(1966)}]{mermin1966}%
  \BibitemOpen
  \bibfield  {author} {\bibinfo {author} {\bibfnamefont {N.~D.}\ \bibnamefont
  {Mermin}}\ and\ \bibinfo {author} {\bibfnamefont {H.}~\bibnamefont
  {Wagner}},\ }\href@noop {} {\bibfield  {journal} {\bibinfo  {journal} {Phys.
  Rev. Lett.}\ }\textbf {\bibinfo {volume} {17}},\ \bibinfo {pages} {1133}
  (\bibinfo {year} {1966})}\BibitemShut {NoStop}%
\bibitem [{\citenamefont {Hope}\ \emph {et~al.}(2000)\citenamefont {Hope},
  \citenamefont {Choi}, \citenamefont {Bode},\ and\ \citenamefont
  {Bland}}]{hope2000}%
  \BibitemOpen
  \bibfield  {author} {\bibinfo {author} {\bibfnamefont {S.}~\bibnamefont
  {Hope}}, \bibinfo {author} {\bibfnamefont {B.-C.}\ \bibnamefont {Choi}},
  \bibinfo {author} {\bibfnamefont {P.}~\bibnamefont {Bode}}, \ and\ \bibinfo
  {author} {\bibfnamefont {J.}~\bibnamefont {Bland}},\ }\href@noop {}
  {\bibfield  {journal} {\bibinfo  {journal} {Phys. Rev. B}\ }\textbf {\bibinfo
  {volume} {61}},\ \bibinfo {pages} {5876} (\bibinfo {year}
  {2000})}\BibitemShut {NoStop}%
\bibitem [{\citenamefont {Rasmussen}\ and\ \citenamefont
  {Thygesen}(2015)}]{rasmussen2015}%
  \BibitemOpen
  \bibfield  {author} {\bibinfo {author} {\bibfnamefont {F.~A.}\ \bibnamefont
  {Rasmussen}}\ and\ \bibinfo {author} {\bibfnamefont {K.~S.}\ \bibnamefont
  {Thygesen}},\ }\href@noop {} {\bibfield  {journal} {\bibinfo  {journal} {J.
  Phys. Chem. C}\ }\textbf {\bibinfo {volume} {119}},\ \bibinfo {pages} {13169}
  (\bibinfo {year} {2015})}\BibitemShut {NoStop}%
\bibitem [{\citenamefont {Rupp}\ \emph {et~al.}(2012)\citenamefont {Rupp},
  \citenamefont {Tkatchenko}, \citenamefont {M\"uller},\ and\ \citenamefont
  {von Lilienfeld}}]{rupp2012}%
  \BibitemOpen
  \bibfield  {author} {\bibinfo {author} {\bibfnamefont {M.}~\bibnamefont
  {Rupp}}, \bibinfo {author} {\bibfnamefont {A.}~\bibnamefont {Tkatchenko}},
  \bibinfo {author} {\bibfnamefont {K.-R.}\ \bibnamefont {M\"uller}}, \ and\
  \bibinfo {author} {\bibfnamefont {O.~A.}\ \bibnamefont {von Lilienfeld}},\
  }\href {\doibase 10.1103/PhysRevLett.108.058301} {\bibfield  {journal}
  {\bibinfo  {journal} {Phys. Rev. Lett.}\ }\textbf {\bibinfo {volume} {108}},\
  \bibinfo {pages} {058301} (\bibinfo {year} {2012})}\BibitemShut {NoStop}%
\bibitem [{\citenamefont {Meredig}\ \emph {et~al.}(2014)\citenamefont
  {Meredig}, \citenamefont {Agrawal}, \citenamefont {Kirklin}, \citenamefont
  {Saal}, \citenamefont {Doak}, \citenamefont {Thompson}, \citenamefont
  {Zhang}, \citenamefont {Choudhary},\ and\ \citenamefont
  {Wolverton}}]{meredig2014}%
  \BibitemOpen
  \bibfield  {author} {\bibinfo {author} {\bibfnamefont {B.}~\bibnamefont
  {Meredig}}, \bibinfo {author} {\bibfnamefont {A.}~\bibnamefont {Agrawal}},
  \bibinfo {author} {\bibfnamefont {S.}~\bibnamefont {Kirklin}}, \bibinfo
  {author} {\bibfnamefont {J.~E.}\ \bibnamefont {Saal}}, \bibinfo {author}
  {\bibfnamefont {J.~W.}\ \bibnamefont {Doak}}, \bibinfo {author}
  {\bibfnamefont {A.}~\bibnamefont {Thompson}}, \bibinfo {author}
  {\bibfnamefont {K.}~\bibnamefont {Zhang}}, \bibinfo {author} {\bibfnamefont
  {A.}~\bibnamefont {Choudhary}}, \ and\ \bibinfo {author} {\bibfnamefont
  {C.}~\bibnamefont {Wolverton}},\ }\href {\doibase 10.1103/PhysRevB.89.094104}
  {\bibfield  {journal} {\bibinfo  {journal} {Phys. Rev. B}\ }\textbf {\bibinfo
  {volume} {89}},\ \bibinfo {pages} {094104} (\bibinfo {year}
  {2014})}\BibitemShut {NoStop}%
\bibitem [{\citenamefont {Seko}\ \emph {et~al.}(2015)\citenamefont {Seko},
  \citenamefont {Togo}, \citenamefont {Hayashi}, \citenamefont {Tsuda},
  \citenamefont {Chaput},\ and\ \citenamefont {Tanaka}}]{tsuda2015}%
  \BibitemOpen
  \bibfield  {author} {\bibinfo {author} {\bibfnamefont {A.}~\bibnamefont
  {Seko}}, \bibinfo {author} {\bibfnamefont {A.}~\bibnamefont {Togo}}, \bibinfo
  {author} {\bibfnamefont {H.}~\bibnamefont {Hayashi}}, \bibinfo {author}
  {\bibfnamefont {K.}~\bibnamefont {Tsuda}}, \bibinfo {author} {\bibfnamefont
  {L.}~\bibnamefont {Chaput}}, \ and\ \bibinfo {author} {\bibfnamefont
  {I.}~\bibnamefont {Tanaka}},\ }\href {\doibase
  10.1103/PhysRevLett.115.205901} {\bibfield  {journal} {\bibinfo  {journal}
  {Phys. Rev. Lett.}\ }\textbf {\bibinfo {volume} {115}},\ \bibinfo {pages}
  {205901} (\bibinfo {year} {2015})}\BibitemShut {NoStop}%
\bibitem [{\citenamefont {Ueno}\ \emph {et~al.}(2016)\citenamefont {Ueno},
  \citenamefont {Rhone}, \citenamefont {Hou}, \citenamefont {Mizoguchi},\ and\
  \citenamefont {Tsuda}}]{Ueno2016}%
  \BibitemOpen
  \bibfield  {author} {\bibinfo {author} {\bibfnamefont {T.}~\bibnamefont
  {Ueno}}, \bibinfo {author} {\bibfnamefont {T.~D.}\ \bibnamefont {Rhone}},
  \bibinfo {author} {\bibfnamefont {Z.}~\bibnamefont {Hou}}, \bibinfo {author}
  {\bibfnamefont {T.}~\bibnamefont {Mizoguchi}}, \ and\ \bibinfo {author}
  {\bibfnamefont {K.}~\bibnamefont {Tsuda}},\ }\href@noop {} {\bibfield
  {journal} {\bibinfo  {journal} {Materials Discovery}\ }\textbf {\bibinfo
  {volume} {4}},\ \bibinfo {pages} {18} (\bibinfo {year} {2016})}\BibitemShut
  {NoStop}%
\bibitem [{\citenamefont {Choudhary}\ \emph {et~al.}(2017)\citenamefont
  {Choudhary}, \citenamefont {Kalish}, \citenamefont {Beams},\ and\
  \citenamefont {Tavazza}}]{Francesca2017}%
  \BibitemOpen
  \bibfield  {author} {\bibinfo {author} {\bibfnamefont {K.}~\bibnamefont
  {Choudhary}}, \bibinfo {author} {\bibfnamefont {I.}~\bibnamefont {Kalish}},
  \bibinfo {author} {\bibfnamefont {R.}~\bibnamefont {Beams}}, \ and\ \bibinfo
  {author} {\bibfnamefont {F.}~\bibnamefont {Tavazza}},\ }\href@noop {}
  {\bibfield  {journal} {\bibinfo  {journal} {Sci. Rep.}\ }\textbf {\bibinfo
  {volume} {7}},\ \bibinfo {pages} {5179} (\bibinfo {year} {2017})}\BibitemShut
  {NoStop}%
\bibitem [{\citenamefont {Ju}\ \emph {et~al.}(2017)\citenamefont {Ju},
  \citenamefont {Shiga}, \citenamefont {Feng}, \citenamefont {Hou},
  \citenamefont {Tsuda},\ and\ \citenamefont {Shiomi}}]{tsuda2017}%
  \BibitemOpen
  \bibfield  {author} {\bibinfo {author} {\bibfnamefont {S.}~\bibnamefont
  {Ju}}, \bibinfo {author} {\bibfnamefont {T.}~\bibnamefont {Shiga}}, \bibinfo
  {author} {\bibfnamefont {L.}~\bibnamefont {Feng}}, \bibinfo {author}
  {\bibfnamefont {Z.}~\bibnamefont {Hou}}, \bibinfo {author} {\bibfnamefont
  {K.}~\bibnamefont {Tsuda}}, \ and\ \bibinfo {author} {\bibfnamefont
  {J.}~\bibnamefont {Shiomi}},\ }\href {\doibase 10.1103/PhysRevX.7.021024}
  {\bibfield  {journal} {\bibinfo  {journal} {Phys. Rev. X}\ }\textbf {\bibinfo
  {volume} {7}},\ \bibinfo {pages} {021024} (\bibinfo {year}
  {2017})}\BibitemShut {NoStop}%
\bibitem [{\citenamefont {Landrum}\ and\ \citenamefont
  {Genin}(2003)}]{landrum2003}%
  \BibitemOpen
  \bibfield  {author} {\bibinfo {author} {\bibfnamefont {G.~A.}\ \bibnamefont
  {Landrum}}\ and\ \bibinfo {author} {\bibfnamefont {H.}~\bibnamefont
  {Genin}},\ }\href@noop {} {\bibfield  {journal} {\bibinfo  {journal} {J.
  Solid State Chem.}\ }\textbf {\bibinfo {volume} {176}},\ \bibinfo {pages}
  {587} (\bibinfo {year} {2003})}\BibitemShut {NoStop}%
\bibitem [{\citenamefont {Lu}\ \emph {et~al.}(2017)\citenamefont {Lu},
  \citenamefont {Zhu}, \citenamefont {Xiao}, \citenamefont {Chuu},
  \citenamefont {Han}, \citenamefont {Chiu}, \citenamefont {Cheng},
  \citenamefont {Yang}, \citenamefont {Wei}, \citenamefont {Yang},
  \citenamefont {Wang}, \citenamefont {Sokaras}, \citenamefont {Nordlund},
  \citenamefont {Yang}, \citenamefont {Muller}, \citenamefont {Chou},
  \citenamefont {Zhang},\ and\ \citenamefont {Li}}]{janus2017}%
  \BibitemOpen
  \bibfield  {author} {\bibinfo {author} {\bibfnamefont {A.-Y.}\ \bibnamefont
  {Lu}}, \bibinfo {author} {\bibfnamefont {H.}~\bibnamefont {Zhu}}, \bibinfo
  {author} {\bibfnamefont {J.}~\bibnamefont {Xiao}}, \bibinfo {author}
  {\bibfnamefont {C.-P.}\ \bibnamefont {Chuu}}, \bibinfo {author}
  {\bibfnamefont {Y.}~\bibnamefont {Han}}, \bibinfo {author} {\bibfnamefont
  {M.-H.}\ \bibnamefont {Chiu}}, \bibinfo {author} {\bibfnamefont {C.-C.}\
  \bibnamefont {Cheng}}, \bibinfo {author} {\bibfnamefont {C.-W.}\ \bibnamefont
  {Yang}}, \bibinfo {author} {\bibfnamefont {K.-H.}\ \bibnamefont {Wei}},
  \bibinfo {author} {\bibfnamefont {Y.}~\bibnamefont {Yang}}, \bibinfo {author}
  {\bibfnamefont {Y.}~\bibnamefont {Wang}}, \bibinfo {author} {\bibfnamefont
  {D.}~\bibnamefont {Sokaras}}, \bibinfo {author} {\bibfnamefont
  {D.}~\bibnamefont {Nordlund}}, \bibinfo {author} {\bibfnamefont
  {P.}~\bibnamefont {Yang}}, \bibinfo {author} {\bibfnamefont {D.~A.}\
  \bibnamefont {Muller}}, \bibinfo {author} {\bibfnamefont {M.-Y.}\
  \bibnamefont {Chou}}, \bibinfo {author} {\bibfnamefont {X.}~\bibnamefont
  {Zhang}}, \ and\ \bibinfo {author} {\bibfnamefont {L.-J.}\ \bibnamefont
  {Li}},\ }\href@noop {} {\bibfield  {journal} {\bibinfo  {journal} {Nat.
  Nanotechnol.}\ }\textbf {\bibinfo {volume} {12}},\ \bibinfo {pages} {744}
  (\bibinfo {year} {2017})}\BibitemShut {NoStop}%
\bibitem [{\citenamefont {Schoenholz}\ \emph {et~al.}(2016)\citenamefont
  {Schoenholz}, \citenamefont {Cubuk}, \citenamefont {Sussman}, \citenamefont
  {Kaxiras},\ and\ \citenamefont {Liu}}]{cubuk2016}%
  \BibitemOpen
  \bibfield  {author} {\bibinfo {author} {\bibfnamefont {S.~S.}\ \bibnamefont
  {Schoenholz}}, \bibinfo {author} {\bibfnamefont {E.~D.}\ \bibnamefont
  {Cubuk}}, \bibinfo {author} {\bibfnamefont {D.~M.}\ \bibnamefont {Sussman}},
  \bibinfo {author} {\bibfnamefont {E.}~\bibnamefont {Kaxiras}}, \ and\
  \bibinfo {author} {\bibfnamefont {A.~J.}\ \bibnamefont {Liu}},\ }\href@noop
  {} {\bibfield  {journal} {\bibinfo  {journal} {Nat. Phys.}\ }\textbf
  {\bibinfo {volume} {12}},\ \bibinfo {pages} {469} (\bibinfo {year}
  {2016})}\BibitemShut {NoStop}%
\bibitem [{\citenamefont {Cubuk}\ \emph {et~al.}(2017)\citenamefont {Cubuk},
  \citenamefont {Malone}, \citenamefont {Onat}, \citenamefont {Waterland},\
  and\ \citenamefont {Kaxiras}}]{cubuk2017}%
  \BibitemOpen
  \bibfield  {author} {\bibinfo {author} {\bibfnamefont {E.~D.}\ \bibnamefont
  {Cubuk}}, \bibinfo {author} {\bibfnamefont {B.~D.}\ \bibnamefont {Malone}},
  \bibinfo {author} {\bibfnamefont {B.}~\bibnamefont {Onat}}, \bibinfo {author}
  {\bibfnamefont {A.}~\bibnamefont {Waterland}}, \ and\ \bibinfo {author}
  {\bibfnamefont {E.}~\bibnamefont {Kaxiras}},\ }\href@noop {} {\bibfield
  {journal} {\bibinfo  {journal} {J. Chem. Phys.}\ }\textbf {\bibinfo {volume}
  {147}},\ \bibinfo {pages} {024104} (\bibinfo {year} {2017})}\BibitemShut
  {NoStop}%
\bibitem [{Note1()}]{Note1}%
  \BibitemOpen
  \bibinfo {note} {The results of these DFT calculations will be used to build
  a database of monolayer 2D materials which will be publicly available to the
  scientific community.}\BibitemShut {Stop}%
\bibitem [{\citenamefont {Heinz}\ and\ \citenamefont
  {Suter}(2004)}]{Ulrich2004}%
  \BibitemOpen
  \bibfield  {author} {\bibinfo {author} {\bibfnamefont {H.}~\bibnamefont
  {Heinz}}\ and\ \bibinfo {author} {\bibfnamefont {U.~W.}\ \bibnamefont
  {Suter}},\ }\href@noop {} {\bibfield  {journal} {\bibinfo  {journal} {J.
  Phys. Chem. B}\ }\textbf {\bibinfo {volume} {108}},\ \bibinfo {pages} {18341}
  (\bibinfo {year} {2004})}\BibitemShut {NoStop}%
\bibitem [{Note2()}]{Note2}%
  \BibitemOpen
  \bibinfo {note} {We used the GGA-PBE for the exchange-correlation functional.
  The energy cutoff was 300 eV. The vacuum region was thicker than 20 \r A. The
  atoms were fully relaxed until the force on each atom was smaller than 0.01
  eV/\r A. A $\Gamma $-centered 10$\times $10$\times $1 \protect \emph
  {k}-point mesh was utilized.}\BibitemShut {Stop}%
\bibitem [{\citenamefont {Kresse}\ and\ \citenamefont
  {Furthm\"uller}(1996)}]{vasp3}%
  \BibitemOpen
  \bibfield  {author} {\bibinfo {author} {\bibfnamefont {G.}~\bibnamefont
  {Kresse}}\ and\ \bibinfo {author} {\bibfnamefont {J.}~\bibnamefont
  {Furthm\"uller}},\ }\href {\doibase 10.1103/PhysRevB.54.11169} {\bibfield
  {journal} {\bibinfo  {journal} {Phys. Rev. B}\ }\textbf {\bibinfo {volume}
  {54}},\ \bibinfo {pages} {11169} (\bibinfo {year} {1996})}\BibitemShut
  {NoStop}%
\bibitem [{\citenamefont {Ghiringhelli}\ \emph {et~al.}(2015)\citenamefont
  {Ghiringhelli}, \citenamefont {Vybiral}, \citenamefont {Levchenko},
  \citenamefont {Draxl},\ and\ \citenamefont {Scheffler}}]{Scheffler2015}%
  \BibitemOpen
  \bibfield  {author} {\bibinfo {author} {\bibfnamefont {L.~M.}\ \bibnamefont
  {Ghiringhelli}}, \bibinfo {author} {\bibfnamefont {J.}~\bibnamefont
  {Vybiral}}, \bibinfo {author} {\bibfnamefont {S.~V.}\ \bibnamefont
  {Levchenko}}, \bibinfo {author} {\bibfnamefont {C.}~\bibnamefont {Draxl}}, \
  and\ \bibinfo {author} {\bibfnamefont {M.}~\bibnamefont {Scheffler}},\ }\href
  {\doibase 10.1103/PhysRevLett.114.105503} {\bibfield  {journal} {\bibinfo
  {journal} {Phys. Rev. Lett.}\ }\textbf {\bibinfo {volume} {114}},\ \bibinfo
  {pages} {105503} (\bibinfo {year} {2015})}\BibitemShut {NoStop}%
\bibitem [{\citenamefont {Seko}\ \emph {et~al.}(2017)\citenamefont {Seko},
  \citenamefont {Hayashi}, \citenamefont {Nakayama}, \citenamefont
  {Takahashi},\ and\ \citenamefont {Tanaka}}]{Tanaka2017}%
  \BibitemOpen
  \bibfield  {author} {\bibinfo {author} {\bibfnamefont {A.}~\bibnamefont
  {Seko}}, \bibinfo {author} {\bibfnamefont {H.}~\bibnamefont {Hayashi}},
  \bibinfo {author} {\bibfnamefont {K.}~\bibnamefont {Nakayama}}, \bibinfo
  {author} {\bibfnamefont {A.}~\bibnamefont {Takahashi}}, \ and\ \bibinfo
  {author} {\bibfnamefont {I.}~\bibnamefont {Tanaka}},\ }\href {\doibase
  10.1103/PhysRevB.95.144110} {\bibfield  {journal} {\bibinfo  {journal} {Phys.
  Rev. B}\ }\textbf {\bibinfo {volume} {95}},\ \bibinfo {pages} {144110}
  (\bibinfo {year} {2017})}\BibitemShut {NoStop}%
\bibitem [{\citenamefont {Mentel}(2014)}]{mendeleev2014}%
  \BibitemOpen
  \bibfield  {author} {\bibinfo {author} {\bibfnamefont {L.}~\bibnamefont
  {Mentel}},\ }\href {https://bitbucket.org/lukaszmentel/mendeleev} {\enquote
  {\bibinfo {title} {{Mendeleev} -- a python resource for properties of
  chemical elements, ions and isotopes},}\ } (\bibinfo {year} {2014}),\
  \bibinfo {note} {0.4.1}\BibitemShut {NoStop}%
\bibitem [{\citenamefont {Chikazumi}\ and\ \citenamefont
  {Graham}(2009)}]{chikazumi}%
  \BibitemOpen
  \bibfield  {author} {\bibinfo {author} {\bibfnamefont {S.}~\bibnamefont
  {Chikazumi}}\ and\ \bibinfo {author} {\bibfnamefont {C.~D.}\ \bibnamefont
  {Graham}},\ }\href@noop {} {\emph {\bibinfo {title} {Physics of
  Ferromagnetism}}},\ Vol.~\bibinfo {volume} {94}\ (\bibinfo  {publisher}
  {Oxford University Press on Demand},\ \bibinfo {year} {2009})\BibitemShut
  {NoStop}%
\bibitem [{Note3()}]{Note3}%
  \BibitemOpen
  \bibinfo {note} {See Supplemental Material at http://link.aps.org/
  supplemental/10.1103/PhysRevX.x.xxxxx for more details and discussions
  related to this letter. \label {footnote_sm}}\BibitemShut {NoStop}%
\bibitem [{\citenamefont {Tibshirani}\ \emph {et~al.}(2013)\citenamefont
  {Tibshirani}, \citenamefont {James}, \citenamefont {Witten},\ and\
  \citenamefont {Hastie}}]{hastie}%
  \BibitemOpen
  \bibfield  {author} {\bibinfo {author} {\bibfnamefont {R.}~\bibnamefont
  {Tibshirani}}, \bibinfo {author} {\bibfnamefont {G.}~\bibnamefont {James}},
  \bibinfo {author} {\bibfnamefont {D.}~\bibnamefont {Witten}}, \ and\ \bibinfo
  {author} {\bibfnamefont {T.}~\bibnamefont {Hastie}},\ }\href@noop {}
  {\enquote {\bibinfo {title} {An introduction to statistical learning $-$ with
  applications in {R}},}\ } (\bibinfo {year} {2013})\BibitemShut {NoStop}%
\bibitem [{\citenamefont {Sivadas}\ \emph {et~al.}(2015)\citenamefont
  {Sivadas}, \citenamefont {Daniels}, \citenamefont {Swendsen}, \citenamefont
  {Okamoto},\ and\ \citenamefont {Xiao}}]{sivadas2015}%
  \BibitemOpen
  \bibfield  {author} {\bibinfo {author} {\bibfnamefont {N.}~\bibnamefont
  {Sivadas}}, \bibinfo {author} {\bibfnamefont {M.~W.}\ \bibnamefont
  {Daniels}}, \bibinfo {author} {\bibfnamefont {R.~H.}\ \bibnamefont
  {Swendsen}}, \bibinfo {author} {\bibfnamefont {S.}~\bibnamefont {Okamoto}}, \
  and\ \bibinfo {author} {\bibfnamefont {D.}~\bibnamefont {Xiao}},\ }\href
  {\doibase 10.1103/PhysRevB.91.235425} {\bibfield  {journal} {\bibinfo
  {journal} {Phys. Rev. B}\ }\textbf {\bibinfo {volume} {91}},\ \bibinfo
  {pages} {235425} (\bibinfo {year} {2015})}\BibitemShut {NoStop}%
\bibitem [{\citenamefont {Goldsmith}\ \emph {et~al.}(2017)\citenamefont
  {Goldsmith}, \citenamefont {Boley}, \citenamefont {Vreeken}, \citenamefont
  {Scheffler},\ and\ \citenamefont {Ghiringhelli}}]{scheffler2017}%
  \BibitemOpen
  \bibfield  {author} {\bibinfo {author} {\bibfnamefont {B.~R.}\ \bibnamefont
  {Goldsmith}}, \bibinfo {author} {\bibfnamefont {M.}~\bibnamefont {Boley}},
  \bibinfo {author} {\bibfnamefont {J.}~\bibnamefont {Vreeken}}, \bibinfo
  {author} {\bibfnamefont {M.}~\bibnamefont {Scheffler}}, \ and\ \bibinfo
  {author} {\bibfnamefont {L.~M.}\ \bibnamefont {Ghiringhelli}},\ }\href@noop
  {} {\bibfield  {journal} {\bibinfo  {journal} {New J. Phys.}\ }\textbf
  {\bibinfo {volume} {19}},\ \bibinfo {pages} {013031} (\bibinfo {year}
  {2017})}\BibitemShut {NoStop}%
\bibitem [{\citenamefont {Reshef}\ \emph {et~al.}(2011)\citenamefont {Reshef},
  \citenamefont {Reshef}, \citenamefont {Finucane}, \citenamefont {Grossman},
  \citenamefont {McVean}, \citenamefont {Turnbaugh}, \citenamefont {Lander},
  \citenamefont {Mitzenmacher},\ and\ \citenamefont {Sabeti}}]{reshef}%
  \BibitemOpen
  \bibfield  {author} {\bibinfo {author} {\bibfnamefont {D.~N.}\ \bibnamefont
  {Reshef}}, \bibinfo {author} {\bibfnamefont {Y.~A.}\ \bibnamefont {Reshef}},
  \bibinfo {author} {\bibfnamefont {H.~K.}\ \bibnamefont {Finucane}}, \bibinfo
  {author} {\bibfnamefont {S.~R.}\ \bibnamefont {Grossman}}, \bibinfo {author}
  {\bibfnamefont {G.}~\bibnamefont {McVean}}, \bibinfo {author} {\bibfnamefont
  {P.~J.}\ \bibnamefont {Turnbaugh}}, \bibinfo {author} {\bibfnamefont {E.~S.}\
  \bibnamefont {Lander}}, \bibinfo {author} {\bibfnamefont {M.}~\bibnamefont
  {Mitzenmacher}}, \ and\ \bibinfo {author} {\bibfnamefont {P.~C.}\
  \bibnamefont {Sabeti}},\ }\href@noop {} {\bibfield  {journal} {\bibinfo
  {journal} {Science}\ }\textbf {\bibinfo {volume} {334}},\ \bibinfo {pages}
  {1518} (\bibinfo {year} {2011})}\BibitemShut {NoStop}%
\bibitem [{\citenamefont {Pedregosa}\ \emph {et~al.}(2011)\citenamefont
  {Pedregosa}, \citenamefont {Varoquaux}, \citenamefont {Gramfort},
  \citenamefont {Michel}, \citenamefont {Thirion}, \citenamefont {Grisel},
  \citenamefont {Blondel}, \citenamefont {Prettenhofer}, \citenamefont {Weiss},
  \citenamefont {Dubourg} \emph {et~al.}}]{scikitlearn}%
  \BibitemOpen
  \bibfield  {author} {\bibinfo {author} {\bibfnamefont {F.}~\bibnamefont
  {Pedregosa}}, \bibinfo {author} {\bibfnamefont {G.}~\bibnamefont
  {Varoquaux}}, \bibinfo {author} {\bibfnamefont {A.}~\bibnamefont {Gramfort}},
  \bibinfo {author} {\bibfnamefont {V.}~\bibnamefont {Michel}}, \bibinfo
  {author} {\bibfnamefont {B.}~\bibnamefont {Thirion}}, \bibinfo {author}
  {\bibfnamefont {O.}~\bibnamefont {Grisel}}, \bibinfo {author} {\bibfnamefont
  {M.}~\bibnamefont {Blondel}}, \bibinfo {author} {\bibfnamefont
  {P.}~\bibnamefont {Prettenhofer}}, \bibinfo {author} {\bibfnamefont
  {R.}~\bibnamefont {Weiss}}, \bibinfo {author} {\bibfnamefont
  {V.}~\bibnamefont {Dubourg}},  \emph {et~al.},\ }\href@noop {} {\bibfield
  {journal} {\bibinfo  {journal} {J. Mach. Learn. Res.}\ }\textbf {\bibinfo
  {volume} {12}},\ \bibinfo {pages} {2825} (\bibinfo {year}
  {2011})}\BibitemShut {NoStop}%
\bibitem [{Note4()}]{Note4}%
  \BibitemOpen
  \bibinfo {note} {The deep neural network used in this study is implemented by
  tensorflow. It is comprised of 3 hidden layers with sizes 10, 30 and 10
  units}\BibitemShut {NoStop}%
\bibitem [{\citenamefont {Han}(2016)}]{han2016}%
  \BibitemOpen
  \bibfield  {author} {\bibinfo {author} {\bibfnamefont {W.}~\bibnamefont
  {Han}},\ }\href@noop {} {\bibfield  {journal} {\bibinfo  {journal} {APL
  Mater.}\ }\textbf {\bibinfo {volume} {4}},\ \bibinfo {pages} {032401}
  (\bibinfo {year} {2016})}\BibitemShut {NoStop}%
\bibitem [{\citenamefont {Chen}\ \emph
  {et~al.}(2013{\natexlab{b}})\citenamefont {Chen}, \citenamefont {Yang},
  \citenamefont {Wang}, \citenamefont {Imai}, \citenamefont {Ohta},
  \citenamefont {Michioka}, \citenamefont {Yoshimura},\ and\ \citenamefont
  {Fang}}]{chen2013}%
  \BibitemOpen
  \bibfield  {author} {\bibinfo {author} {\bibfnamefont {B.}~\bibnamefont
  {Chen}}, \bibinfo {author} {\bibfnamefont {J.}~\bibnamefont {Yang}}, \bibinfo
  {author} {\bibfnamefont {H.}~\bibnamefont {Wang}}, \bibinfo {author}
  {\bibfnamefont {M.}~\bibnamefont {Imai}}, \bibinfo {author} {\bibfnamefont
  {H.}~\bibnamefont {Ohta}}, \bibinfo {author} {\bibfnamefont {C.}~\bibnamefont
  {Michioka}}, \bibinfo {author} {\bibfnamefont {K.}~\bibnamefont {Yoshimura}},
  \ and\ \bibinfo {author} {\bibfnamefont {M.}~\bibnamefont {Fang}},\
  }\href@noop {} {\bibfield  {journal} {\bibinfo  {journal} {J. Phys. Soc.
  Jpn}\ }\textbf {\bibinfo {volume} {82}},\ \bibinfo {pages} {124711} (\bibinfo
  {year} {2013}{\natexlab{b}})}\BibitemShut {NoStop}%
\end{thebibliography}

%

\end{document}